\documentclass{aa}  
\usepackage{graphicx}
\usepackage[export]{adjustbox}
\usepackage{txfonts}
\usepackage[colorlinks, linktoc=all]{hyperref}
\hypersetup{linkcolor=blue,citecolor=black,filecolor=black,urlcolor=black}
\usepackage{caption}
\usepackage{amsmath}
\usepackage{subcaption}
\usepackage{blindtext}
\usepackage[normalem]{ulem}
\usepackage[english]{babel}
\usepackage[stable]{footmisc}
\addto\extrasenglish{%
}
\addto\extrasenglish{%
}
\begin{document}

   \title{Influence of migration models and thermal torque on planetary growth in the pebble accretion scenario}
   
   
   \author{Thomas\ Baumann\inst{1}\thanks{\email{Thomas.Baumann@cebaumann.de}}
        \and
        Bertram\ Bitsch\inst{1}
   }
   
   \institute{$^1$Max-Planck-Institut f\"ur Astronomie, K\"onigstuhl 17, 69117 Heidelberg, Germany
   }

  \date{Received January 27, 2020; accepted March 17, 2020}

 
  \abstract
  {
                Low-mass planets that are in the process of growing larger within protoplanetary disks exchange torques with the disk and change their semi-major axis accordingly. 
                This process is called type I migration and is strongly dependent on the underlying disk structure.
                As a result, there are many uncertainties about planetary migration in general.
                In a number of simulations, the current type I migration rates lead to planets reaching the inner edge of the disk within the disk lifetime.
                A new kind of torque exchange between planet and disk, the thermal torque, aims to slow down inward migration via the heating torque.
                The heating torque may even cause planets to migrate outwards, if the planetary luminosity is large enough.
                Here, we study the influence on planetary migration of the thermal torque on top of previous type I models.
                We find  that the formula of \cite{Paardekooper}  allows for more outward migration than that of \cite{ImprovedTypeI} in most configurations, but we also find that planets evolve to very similar mass and final orbital radius using both formulae in a single planet-formation scenario, including pebble and gas accretion.
                Adding the thermal torque can introduce new, but small, regions of outwards migration if the accretion rates onto the planet correspond to typical solid accretion rates following the pebble accretion scenario.
                If the accretion rates onto the planets become very large, as could be the case in environments with large pebble fluxes (e.g.,\ high-metallicity environments), the thermal torque can allow more efficient outward migration.
                However, even then, the changes for the final mass and orbital positions in our planet formation scenario are quite small.
                This implies that for single planet evolution scenarios, the influence of the heating torque is probably negligible.
  }
  

        \titlerunning{Migration models and the thermal torque in a pebble accretion core growth scenario}
        \authorrunning{Baumann \& Bitsch}
        \maketitle
%

\section{Introduction}
The detection of the first exoplanet around a solar-type star \citep{mayor1995jupiter} has revealed a new class of planets that we do not have in our own solar system: hot Jupiters.
These planets are similar in mass to our own Jupiter, but their orbits around their central star last just shy of a few hours or days. 
However, at positions so close to the star, the sublimation temperatures for silicates can be exceeded, making in-situ formation impossible, which raises the question of how these planets got into this position.

Gravitational interactions between planets and the protoplanetary disk lead to a change in the semi-major axis of the planets, that is, migration.
Spiral density waves, launched at Lindblad resonances, lead to fast inwards migration \citep{WARD1997}. 
Studies in recent years have also found so-called corotation torques \citep{Tanaka_2002, Baruteau_2008, CorotationRegion}, which can slow down migration or foster outwards migration \citep{Paardekooper_Mellema_2006, kley2008migration, kley2009planet} in certain disk conditions and regions (e.g.,\ \citealt{2013A&A...549A.124B}).

Planetary migration is an important ingredient for planet formation simulations \citep{Ida_2004, Mordasini2009, PebbleAccretion, Ndugu2018}.
However, planet formation simulations cannot incorporate 2D or 3D hydrodynamical simulations to accurately calculate the torques over millions of years.
Thus, the planet formation simulations rely on torque formulae derived from hydrodynamical simulations, such as \mbox{\citet{Masset_2010}, \cite{Paardekooper},} or \cite{ImprovedTypeI}.

Two such models of torque formulae that were derived from fits of hydrodynamical simulations are studied and compared in this work.
These are \cite{Paardekooper}, which was derived from a two-dimensional disk, and \cite{ImprovedTypeI}, which is updated research in a three-dimensional disk.
A new development, which is also investigated here, is the thermal torque \citep{2015Natur.520...63B, ThermalTorque}, which can lead to outwards migration due to heat released by the planet.

The migration rate of planets crucially depends on the radial gradients of the gas surface density and the entropy. 
This implies that the migration rates depend significantly on the underlying disk structure and disk model. 
As the planet grows, it can migrate across several ice lines, which would change the composition of the material the planet can accrete \citep{oberg2011spitzer, madhusudhan2017atmospheric, cridland2017composition}.

\cite{PebbleAccretion} presents a study of planet formation in the pebble accretion scenario in the \cite{DiscModel} disk model, with planets migrating according to the \cite{Paardekooper} formula.
One aim of our work here is to determine if the differences between the new formulae from \cite{ImprovedTypeI} are large compared to the migration rates by \cite{Paardekooper}.
Additionally, we investigate the influence of the heating and cooling torque \citep{ThermalTorque} on the final mass and position of planets.
Here, the relevant quantity is the solid accretion rate onto the planet, which releases the heat needed for the heating torque to work \citep{2015Natur.520...63B}.

A previous study by \cite{Guilera} also studied the heating torque in planet formation simulations and found it had a strong influence on planet migration and thus formation.
Here, we expand on their approach and include a self-consistent calculation of the heating torque within the pebble accretion scenario.
In particular, we investigate the influence of the pebble flux on outwards migration triggered by the heating torque.

That said, our results are only applicable to simulations founded on similar assumptions.
A fundamental simplification of this paper is the modeling of single planet evolution only.
N-body simulations (e.g.,\ \citealt{cossou2013making}, \citealt{Izidoro2019}, \citealt{Bitsch2019}), on the other hand, exhibit many additional influence factors which are not captured in our approach.
For example, the trapping in resonances depends crucially on the relative migration speed between the planets.
So if planets suddenly migrate slightly slower or even outwards, trapping in resonances might operate in a different way compared to situations where planets only migrate inwards.

The structure of this paper is as follows: The methods are listed in \autoref{methods_section}, followed by a comparison of the torque formulae from \cite{Paardekooper} and \cite{ImprovedTypeI} in \autoref{pVSjm-section}.
The influence of the thermal torque is analyzed in \autoref{thermal-section}.
Finally, the analysis of the previous sections is repeated in the disk model from \cite{IdaDisk} in \autoref{idaDisk-section}.
A discussion of the results and the conclusion can be found in \autoref{discussion_section} and \autoref{conclusion_section}.

\section{Planetary growth and migration models}
\label{methods_section}
The model of planetary growth implemented here has already been described in \cite{PebbleAccretion}, so we simply repeat the essentials of the growth model.
We also use the disk model of \cite{DiscModel}, which originates from 3D hydrodynamical simulations, including stellar and viscous heating, as well as radiative cooling by micrometer-sized dust grains. 

In this paper, we assume planetary cores to grow via pebble accretion and migrate in type I and type II fashion.
Pebbles are rocky or icy objects of mm to cm size that form by coagulation or condensation of dust particles \citep{brauer2008planetesimal, birnstiel2012simple, ros2013ice, drkazkowska2017planetesimal, ormel2017formation}.
By accreting these pebbles, planets can grow cores efficiently \citep{ormel2010effect, johansen2010prograde, lambrechts2012astronomy}, that is,\ on timescales that are multiple orders of magnitude shorter than planetesimal accretion \citep{johansen2017forming, johansen2019exploring}.
The size of the pebbles relates to the Stokes number (e.g.,\ \citealt{brauer2008planetesimal}), which is set to 0.1 in these simulations.

In our simulations, planets are inserted at pebble transition mass, meaning they accrete pebbles within the Hill radius right away. 
Pebble transition mass is determined following \cite{lambrechts2012astronomy, lambrechts2014forming}.

Pebble accretion is modeled in two regimes.
The planet accretes pebbles in the fast 2D scenario, when its Hill radius is larger than the pebble scale height,
\begin{equation}
H_{\mathrm{peb}}=H_{\mathrm{gas}} \sqrt{\alpha / \tau_{\mathrm{f}}},
\label{pebble_scale_height}
\end{equation}
with the gas scale height $H_{\mathrm{gas}}$, the $\alpha$-viscosity parameter \citep{alphavisc} and the Stokes number $\tau_{\mathrm{f}}$ \citep{youdin2007particle, 2015Icar..258..418M}.
The accretion rate in the 2D regime is given by:
\begin{equation}
\dot{M}_{\mathrm{c}, 2 \mathrm{D}}=2 r_{\mathrm{H}} v_{\mathrm{H}} \Sigma_{\mathrm{peb}}.
\label{2D_pebble_accretion-eq}
\end{equation}
Here, $r_{\mathrm{H}}$ is the Hill radius, $v_{\mathrm{H}} = r_{\mathrm{H}} \Omega$ and $ \Sigma_{\mathrm{peb}}$ is the pebble surface density at the planet's location.
Otherwise, the planet accretes in the 3D scenario.
A reduction factor from \mbox{\cite{2015Icar..258..418M}} is applied, resulting in an accretion rate of:\begin{equation}
\dot{M}_{\mathrm{c}, 3 \mathrm{D}}=\dot{M}_{\mathrm{c}, 2 \mathrm{D}}\left(\frac{\pi\left(\tau_{\mathrm{f}} / 0.1\right)^{1 / 3} r_{\mathrm{H}}}{2 \sqrt{2 \pi} H_{\mathrm{peb}}}\right).
\label{3D_pebble_accretion-eq}
\end{equation}

The amount of pebbles that are available to the planet is determined by the pebble flux through the disk, which is set to $\dot{M}_\mathrm{peb} = 2 \cdot 10^{-4} \cdot\exp (-\frac{t}{t_\mathrm{f}}) \, \frac{M_\mathrm{E}}{\mathrm{year}}$ in most simulations.
This corresponds to a total of about 380\,$M_\mathrm{E}$ (Earth-masses) of pebbles in the disk throughout its lifetime of 3\,Myr, equivalent to roughly a 1\% dust-to-gas ratio if the initial disk mass was 10\% of the solar mass.

During core growth, the planet accretes 90\% in solids and 10\% in gas, building a gaseous envelope alongside the core.
When the planet reaches pebble isolation mass \citep{lambrechts2014separating, bitsch2018pebble, ataiee2018much}, core growth is halted and the gaseous envelope is accreted onto the planet.
Following \cite{bitsch2018b}, the pebble isolation mass is calculated via: 
\begin{gather}
M_{\mathrm{iso}}=25 f_{\mathrm{fit}}\, \mathrm{M}_{\mathrm{E}}+\frac{\Pi_{\mathrm{crit}}}{\lambda}\, \mathrm{M}_{\mathrm{E}},\\
f_{\mathrm{fit}}=\left[\frac{H / r}{0.05}\right]^{3}\left[0.34\left(\frac{\log \left(\alpha_{3}\right)}{\log \left(\alpha_{\mathrm{disc}}\right)}\right)^{4}+0.66\right]\left[1-\frac{\frac{\partial \ln P}{\partial \ln r}+2.5}{6}\right],
\end{gather}
with $\lambda \approx 0.00476 / f_{\text {fit }}, \Pi_{\text {crit }}=\frac{\alpha_{\text {disc }}}{2 \tau_{\mathrm{f}}}$ and $\alpha_{3}=0.001$.

After the pebble isolation mass is reached, the heating of the envelope through in-falling pebbles stops and the planetary envelope can contract.
Here, we follow the contraction rates from \cite{piso2014minimum}.
Once the envelope mass is similar to the core mass, runaway gas accretion can start and for this, we follow the rates of \cite{machida2010gas}.

Planetary growth is halted when the planet reaches the inner edge of the disk at 0.1\,AU or when the gas disk dissipates after 3\,Myr.
We stop the growth of planets at 0.1\,AU\ because gas accretion is thought to be inefficient in this region due to strong recycling flows that penetrate deeply into the Hill sphere and prevent gas contraction around young planets (e.g.,\ \citealt{cimerman2017hydrodynamics, lambrechts2017reduced}).

Migration is modeled in two regimes, called type I and II.
Type I migration is a consequence of gravitational interaction between the planet and the protoplanetary disk that it grows in.
The two models for type I migration that are used here are described in \autoref{pVSjm-section}, with the formulae in \autoref{Paardekooper_torque-appendix}, \autoref{JM_torque-appendix} and \autoref{thermal_torque_appendix}.
Type I migration is valid for low mass planets.
As planets grow more massive, they start opening a gap in the disk and transition to type II.
In this regime, migration is determined by stellar accretion and is usually slower than type I and directed towards the star.
For a recent review on type I and type II migration, see \cite{PlanetDiskInteraction} and \cite{baruteau2013planet}.
For information about the calculation of type II migration timescales in this paper, see \autoref{typeII_appendix}.

A simplification in these simulations is that the planets move in circular orbits with vanishing inclinations.
This is a sensible approximation, as eccentricity and inclination are damped quite quickly by the gas disk \citep{BitschKley2010, BitschKley2011}.

\section{Comparison of the torque formulae from \cite{Paardekooper} and \cite{ImprovedTypeI}}
\label{pVSjm-section}
\cite{Paardekooper} and \cite{ImprovedTypeI} both performed hydrodynamical simulations in non-barotropic disks with finite viscosity and constant thermal diffusion to derive their formulae.
The major difference is that \cite{Paardekooper} made two-dimensional simulations, while \cite{ImprovedTypeI} updated their findings based on three-dimensional disks.

Generally, the two formulae agree in the physical principles of the torque exchange, but show different numerical factors.
For example, the saturation mechanisms respond similarly to viscosity and thermal diffusivity changes, but for individual values of viscosity, the behavior can be significantly different.

The \cite{Paardekooper} formula incorporates fewer components.
The  components that they included are: the Lindblad torque, which drives inwards migration due to spiral density waves at Lindblad resonances, and the barotropic and entropy related corotation torques.
The corotation torques typically slow down or even set inwards migration in reverse.
The entropy torque is based on the entropy gradient, that is,\ a combination of the temperature and surface density gradients in the disk.
The second contribution is the barotropic torque, resulting from the surface density gradient.

All of the above components are incorporated in \cite{ImprovedTypeI} as well, although the barotropic torque is called vortensity torque in their formulation.
Additionally, there is the temperature torque as another contribution of the midplane temperature gradient.
Finally, \cite{ImprovedTypeI} introduced the viscous coupling term, a result of viscous production of vortensity at the density jumps at the seperatrices of the horseshoe region.

We ought to keep in mind that these formulae describe type I migration.
Once the planet transitions to type II, its migration is determined by the disk properties and differences in these formulae no longer impact migration.

Type II migration is also an active area of research, where it is currently debated if the type II migration speed follows the viscous flow (e.g.,\ \citealt{durmann2015migration}, \citealt{robert2018toward}) or is a scaling of the type I torques with the gap depth (e.g.,\ \citealt{kanagawa2015formation, kanagawa2018radial, ida2019water, johansen2019planetary}).
Here, we use the viscous speeds for planets in type II migration.

\subsection{Differences in the formulae}
\begin{figure*}
        \centering
        \includegraphics[scale=0.45,valign=t]{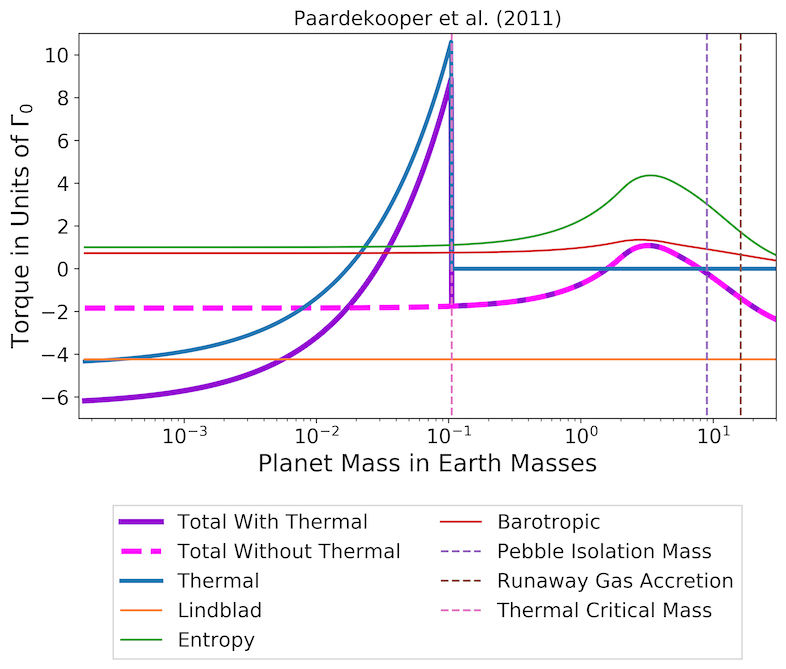}
        \includegraphics[scale=0.45,valign=t]{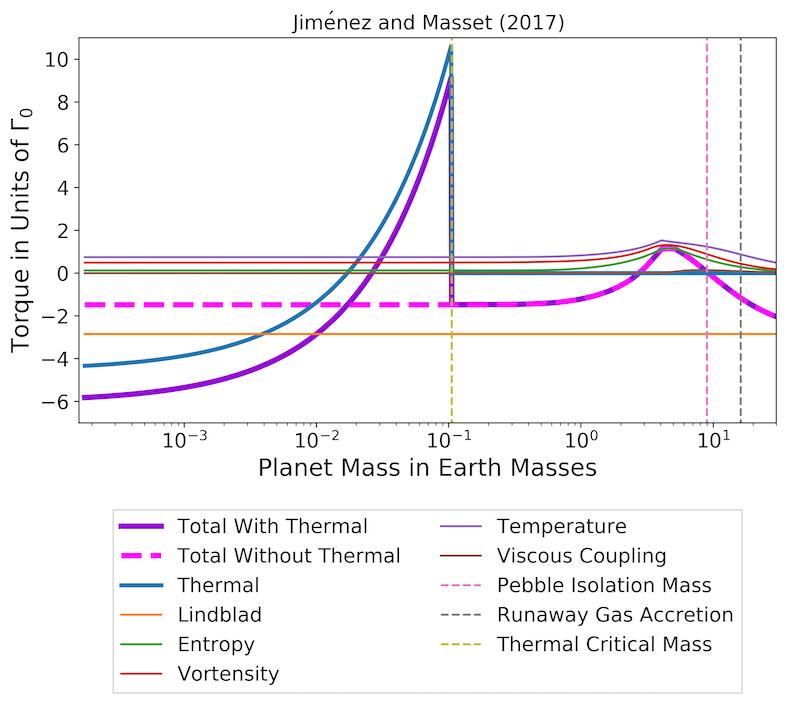}
        \caption{Individual torque components of the \cite{Paardekooper} and \cite{ImprovedTypeI} formulae for planets at \mbox{3\,AU} and 1\,Myr, depending on planetary mass.
                The \cite{ImprovedTypeI} formula adds new contributions, but the total torque looks similar. 
                This is just one arbitrarily chosen set of disk evolution time and position in the disk, which manifests many general trends.
                Outwards migration occurs when the total torque is positive.
                This plot shows outwards migration caused by the heating torque (see \autoref{thermal-section}) for small planets as well as outwards migration caused by the corotation torques for slightly heavier planets.
                The pebble flux through the disk is $\dot{M}_\mathrm{peb} = 2 \cdot 10^{-4} \cdot\exp (-\frac{t}{t_\mathrm{f}}) \, \frac{M_\mathrm{E}}{\mathrm{year}}$, which is relevant for the thermal torque as discussed in \autoref{thermal-section}.
                The viscosity is set to $\alpha = 5.4 \cdot 10^{-3}$ during these simulations.}
        \label{torque_comp-fig}
\end{figure*}

As the formulae are based on the same physics, we expect similar outcomes.
This is true for planet formation, but migration maps can show significant differences.
\autoref{torque_comp-fig} shows the individual torque components of the two formulae for planets of different masses at an orbital radius of 3\,AU and 1\,Myr of disk evolution time.
These plots already include the thermal torque from \cite{ThermalTorque}, which is discussed in \autoref{thermal-section}.
The plot is an arbitrary example of disk evolution time and position in the disk, but serves to show many general trends.

The torques shown in \autoref{torque_comp-fig} are normalized to: 
\begin{equation}
\label{Gamma_0-eq}
\Gamma_0 = \Big ( \frac{q}{h} \Big )^2 \Sigma_\mathrm{P} r^4_\mathrm{P} \Omega^2_\mathrm{P},
\end{equation}
where the index P indicates that the quantities are evaluated at the planets' position.
In this formulation the Lindblad torque does not depend on the planetary mass, as this dependence is hidden in $\Gamma_0$.
The torque is always negative in both formulae, but we find up to 35\% stronger Lindblad torques with the \cite{Paardekooper} formula, where the largest differences occur in the inner disk.

The barotropic or vortensity torque component is very comparable between the formulae, but the entropy torque shows a large difference. 
It is the main source for outwards migration in the \cite{Paardekooper} formula, but is much smaller in magnitude and often even negative in the \cite{ImprovedTypeI} formula.

The temperature torque in \cite{ImprovedTypeI}, which does not exist in \cite{Paardekooper}, gives an additional positive contribution to the torque.
However, it is usually not large enough to make up for the difference in entropy torques compared to \cite{Paardekooper}.
The viscous coupling term, the other new component in the \cite{ImprovedTypeI} formula, is found to be negligible in our simulations.

The horseshoe width, a region where the material moves in horseshoe orbits around the planet, is the source of the corotation torque and is, consequently, an important parameter. 
This width, however, is calculated differently in \cite{ImprovedTypeI} and \cite{Paardekooper}.
In fact, \cite{ImprovedTypeI} have argued that the \cite{Paardekooper} formula is only valid for planets of up to a few Earth-masses.
\cite{ImprovedTypeI} find slightly smaller horseshoe widths at higher masses.
This should lead to smaller corotation torques, but we find that this difference is not that large due to the transition into type II migration for larger planets, which changes the torque in any case.

To summarize, the \cite{Paardekooper} torque formula allows a larger positive contribution from the entropy related corotation torque compared to the \cite{ImprovedTypeI} torque formula.
This leads to larger regions of outwards migration using the \cite{Paardekooper} formula compared to the \cite{ImprovedTypeI} model (see \autoref{migrationmaps_visc-fig}).

\subsection{Impact on planet formation}
\begin{figure*}[!htb]
        \centering
        \begin{minipage}{\linewidth}
                \begin{subfigure}{.49\textwidth}
                        \centering
                        \cite{Paardekooper}
                \end{subfigure}
                \begin{subfigure}{.49\textwidth}
                        \centering
                        \cite{ImprovedTypeI}
                \end{subfigure}
        \end{minipage}
        
        \begin{minipage}{\linewidth}
                \begin{subfigure}{.49\textwidth}
                        \centering
                        \includegraphics[width=\linewidth]{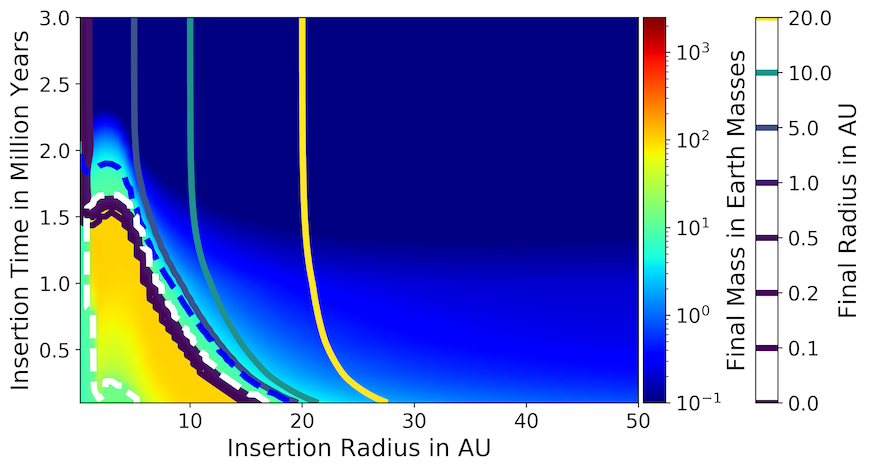}
                \end{subfigure}
                \begin{subfigure}{.49\textwidth}
                        \centering
                        \includegraphics[width=\linewidth]{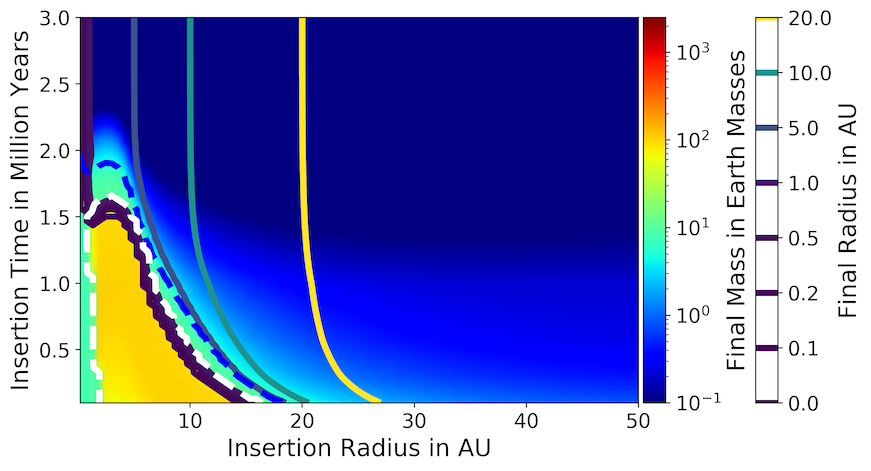}
                \end{subfigure}
                \subcaption{\mbox{$\dot{M}_\mathrm{peb} = 1 \cdot 10^{-4} \cdot\exp (-\frac{t}{t_\mathrm{f}}) \, \frac{M_\mathrm{E}}{\mathrm{year}}$}}
        \end{minipage}
        
        \begin{minipage}{\linewidth}
                \begin{subfigure}{.49\textwidth}
                        \centering
                        \includegraphics[width=\linewidth]{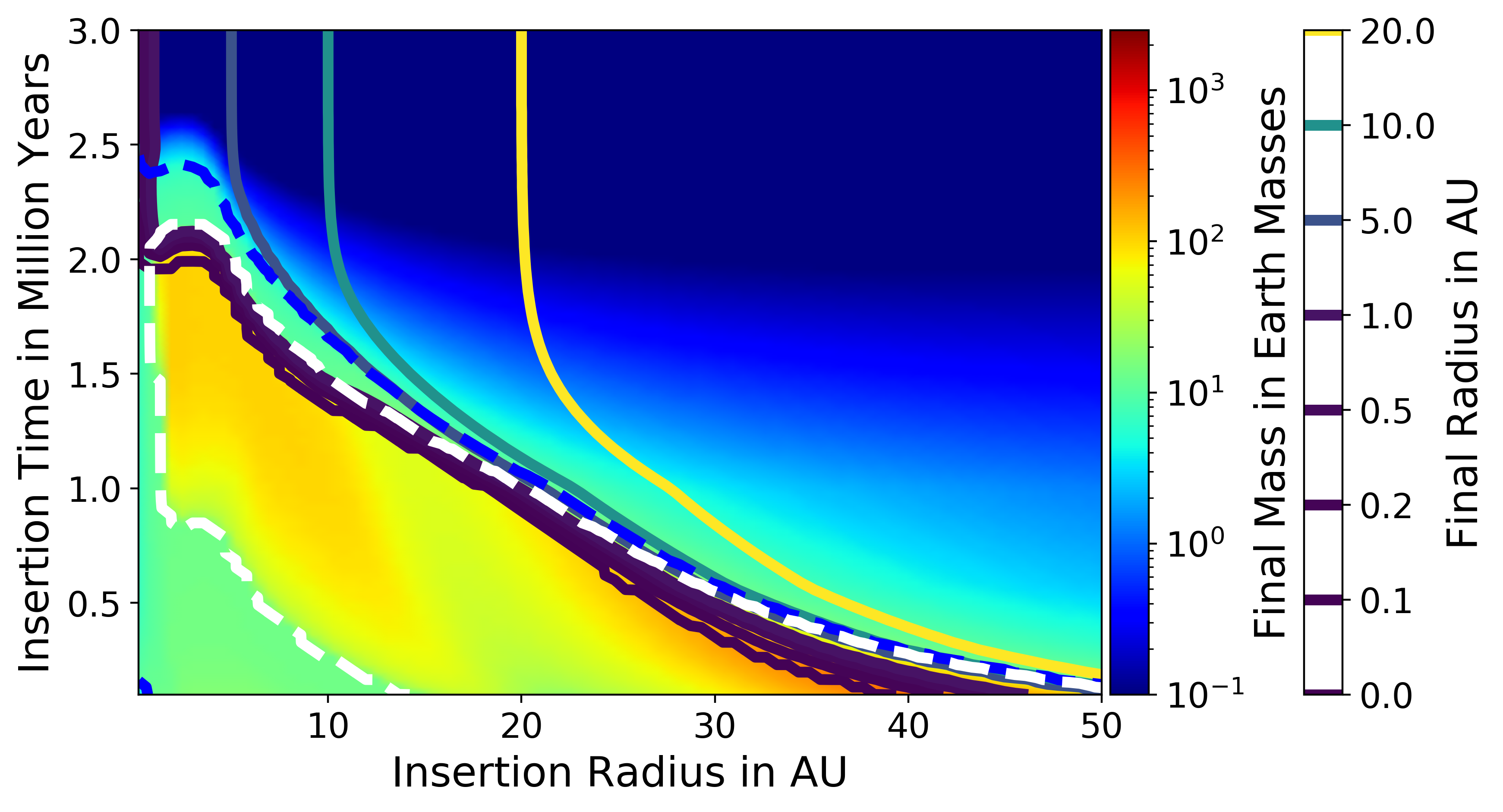}
                \end{subfigure}
                \begin{subfigure}{.49\textwidth}
                        \centering
                        \includegraphics[width=\linewidth]{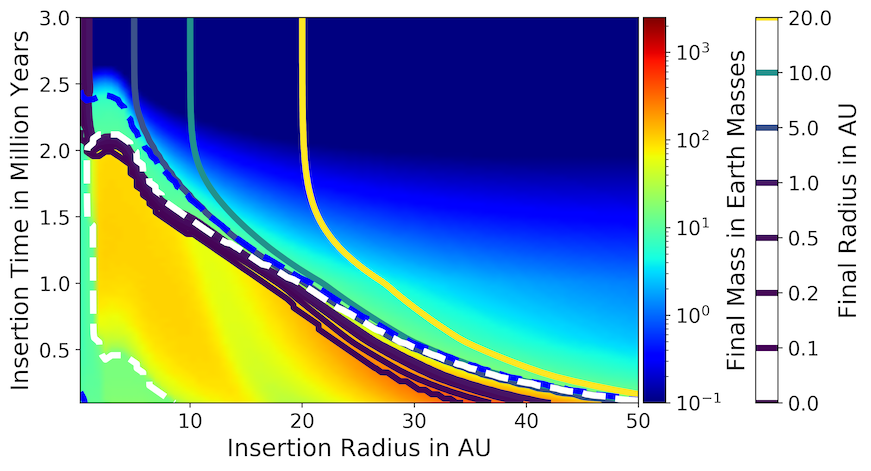}
                \end{subfigure}
                \subcaption{\mbox{$\dot{M}_\mathrm{peb} = 2 \cdot 10^{-4} \cdot\exp (-\frac{t}{t_\mathrm{f}}) \, \frac{M_\mathrm{E}}{\mathrm{year}}$}}
                \label{planet_evo-fig_visc1e-3-fig}
        \end{minipage}
        
        \caption{Planet formation maps with different pebble fluxes in the disk. 
                The axes display where and when a planetary embryo has been inserted in the disk. 
                The color codes how massive the planets grow, while the solid lines outline its final orbital radius.
                Planets that are below the dashed blue line have reached pebble isolation mass and accrete gas. 
                Planets that are interior to the dashed white line have reached runaway gas accretion and can grow very massive. 
                $\dot{M}_\mathrm{peb}$ marks the pebble flux through the disk, which decreases exponentially over time. 
                Differences between the formulae in terms of impact on planet formation are small.
                As the pebble density in the outer disk is small, planets that are inserted there do not grow much.}
        \label{planet_evo-fig}
\end{figure*}

As hinted at before, differences in planet evolution between the two formulae are small.
\autoref{planet_evo-fig} shows the final mass and orbital radius of planets, depending on where ($r_0$) and when ($t_0$) they have been inserted in the disk.
When comparing the four plots, it becomes apparent that the two formulae produce planets of similar mass and final orbital radius, regardless of the amount of pebbles in the disk.
We only observe a difference in the very inner disk, where planets migrating with the \cite{Paardekooper} formula are less massive.
This is caused by the different migration pattern that moves the planets faster into the inner disk, where their pebble isolation mass is small, resulting in smaller planets and thus prevents them from reaching runaway gas accretion.

It is noteworthy that this result only holds for single planet formation.
This is because the migration maps do show significant differences (\autoref{migrationmaps_visc-fig}), which might be of larger influence in an N-body simulation, as they introduce other sources of migration.
The main ones are gravitational interactions between the planets and trapping in mean motion resonances.

The trapping in mean motion resonances depends crucially on the relative migration speed between the planets, where a slower relative migration speed allows trapping in wider resonances compared to faster migration speeds.
The resonant configuration of the system then has important consequences for the stability of the formed planetary system and thus its final configuration (e.g., \citealt{outwardImportance2, Izidoro2019}).

The mutual interactions between the planets could increase the eccentricity of the planets, which quenches the entropy-driven corotation torque and stops outwards migration (e.g.,\ \citealt{BitschKley2010, fendyke2013corotation}).
The impact of forming multiple planets simultaneously is to be tested in future research.

\subsection{Influence of viscosity}
Viscosity has proven to be a large influence factor on migration, because viscosity is responsible to keep the horseshoe region unsaturated (e.g.,\ \citealt{Masset_2001, Baruteau_2008, CorotationRegion, 2013A&A...549A.124B}).
At low viscosity, the corotation torque saturates and outwards migration from the corotation torques ceases to exist.

Our simulations incorporate the $\alpha$-viscosity model from \cite{alphavisc}.
However, we keep the disk structure fixed in our model to show the differences on the torques, even though the disk structure is a combination of viscous and stellar heating as well as radiative cooling \citep{DiscModel}, which would change if the viscosity changes.
When the viscosity was changed in these simulations, only migration timescales and the gap parameter (this parameter models the transition to type II migration, see \autoref{gap_parameter-eq}) were calculated using different values of $\alpha$.
In addition, we assume that there is no influence on the accretion rate by the reduced viscosity, in order to study the effects of lower viscosity on the migration rates separately.

The influence of viscosity is mainly due to two effects.
Type I migration is influenced, as corotation torques saturate more quickly at low viscosity.
This leads to less outwards migration at lower $\alpha$-values.
On the other hand, the gap opening is also changed, meaning planets transition to type II at lower masses for lower viscosity.

\begin{figure*}[!htb]
        \centering
        \begin{minipage}{\linewidth}
                \begin{subfigure}{.49\textwidth}
                        \centering
                        \includegraphics[width=\linewidth]{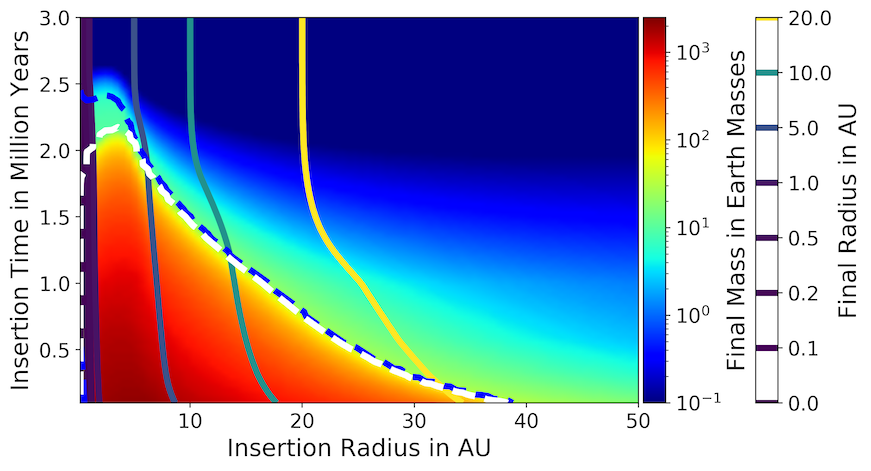}
                        $\alpha = 10^{-4}$
                \end{subfigure}
                \begin{subfigure}{.49\textwidth}
                        \centering
                        \includegraphics[width=\linewidth]{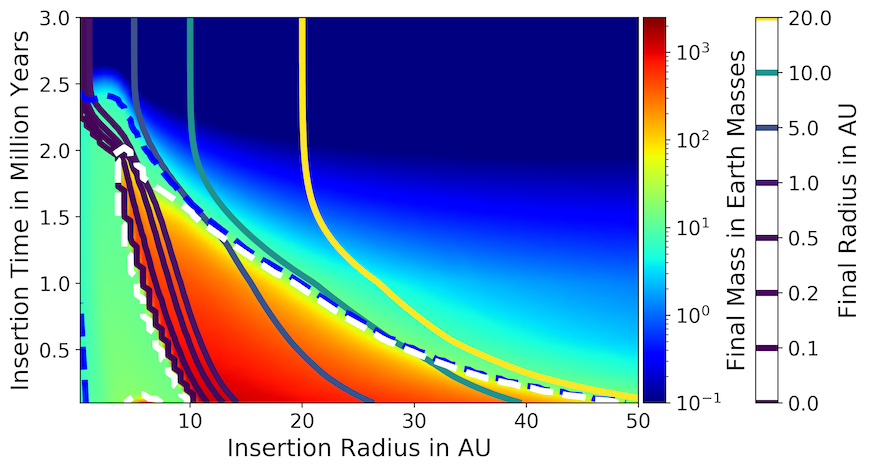}
                        $\alpha = 10^{-3}$
                \end{subfigure}
        \end{minipage}
        
        \begin{minipage}{\linewidth}
                \begin{subfigure}{.49\textwidth}
                        \centering
                        \includegraphics[width=\linewidth]{./figures/_cont_2e-04_100kyrs_JMMC.png}
                        $\alpha = 5.4 \cdot 10^{-3}$ (standard value in our simulations)
                \end{subfigure}
                \begin{subfigure}{.49\textwidth}
                        \centering
                        \includegraphics[width=\linewidth]{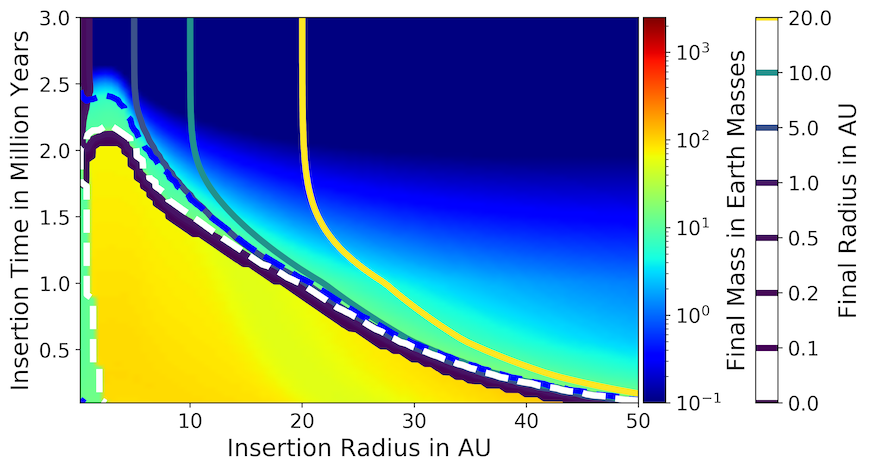}
                        $\alpha = 10^{-2}$
                \end{subfigure}
        \end{minipage}

        \caption{Planet formation maps with different viscosities for planetary migration using the \cite{ImprovedTypeI} formula. 
                Like in \autoref{planet_evo-fig}, planets that are below the dashed blue line have reached pebble isolation mass and accrete gas.
                Planets that are interior to the dashed white line have reached runaway gas accretion and can grow very massive.
                The pebble flux is $\dot{M}_\mathrm{peb} = 2 \cdot 10^{-4} \cdot\exp (-\frac{t}{t_\mathrm{f}}) \, \frac{M_\mathrm{E}}{\mathrm{year}}$ in all plots.
                A lower choice of $\alpha$ mostly leads to more massive planets due to the quicker transition to type II migration.
                In addition, the inwards migration of planets is reduced due to the slow type II migration.}
                \label{visc_evo_fig}
\end{figure*}

The impact of a viscosity change is shown in \autoref{visc_evo_fig}, which shows final mass and orbital radius maps using the \cite{ImprovedTypeI} formula.
Clearly, a choice of lower $\alpha$ leads to more massive planets overall, which can be explained by the earlier transition to type II migration.
The $\alpha = 10^{-3}$ plot also shows planets that do not grow as large when inserted into the inner disk, which is a consequence of the saturation of the corotation torques.
The planets, in this case, grow and migrate to the inner edge of the protoplanetary disk where they stop their migration and growth in our model.
Thus, these planets do not grow into gas giants.

\begin{figure*}[!htb]
        \centering
        \begin{minipage}{\linewidth}
                \centering
                \begin{subfigure}{.44 \textwidth}
                        \centering
                        \cite{Paardekooper}
                \end{subfigure}
                \begin{subfigure}{.44 \textwidth}
                        \centering
                        \cite{ImprovedTypeI}
                \end{subfigure}
        \end{minipage}
        
        \begin{minipage}{\linewidth}
                \centering
                \begin{subfigure}{.44 \textwidth}
                        \centering
                        \includegraphics[width=\linewidth]{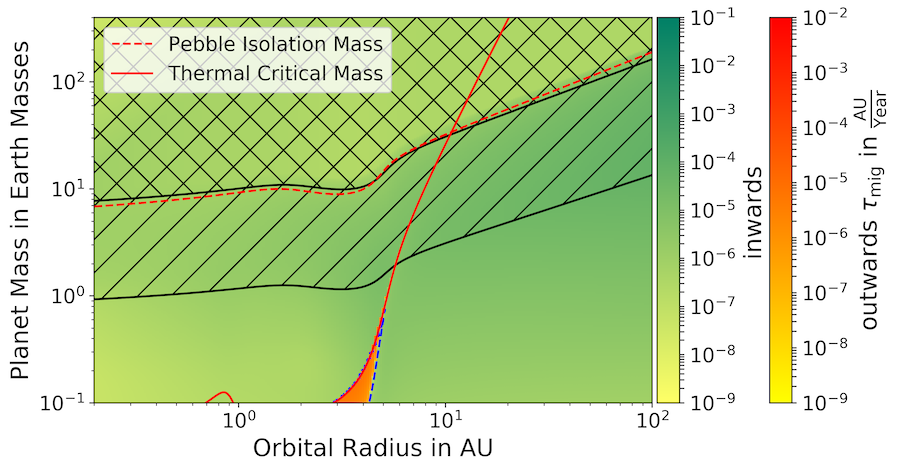}
                \end{subfigure}
                \begin{subfigure}{.44 \textwidth}
                        \centering
                        \includegraphics[width=\linewidth]{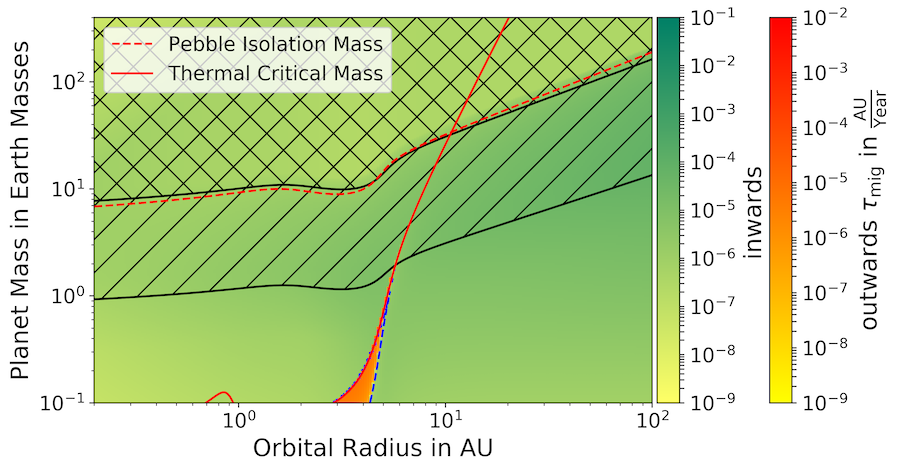}
                \end{subfigure}
                \vskip 0cm
                \subcaption{$\alpha = 10^{-4}$}
                \label{migrationmaps_visc-fig_visc1e-4-fig}
        \end{minipage}
        
        \begin{minipage}{\linewidth}
                \centering
                \begin{subfigure}{.44 \textwidth}
                        \centering
                        \includegraphics[width=\linewidth]{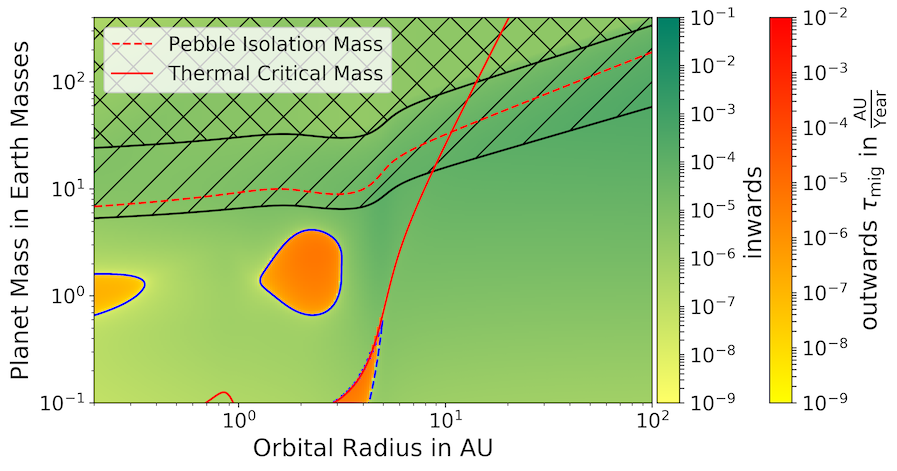}
                \end{subfigure}
                \begin{subfigure}{.44 \textwidth}
                        \centering
                        \includegraphics[width=\linewidth]{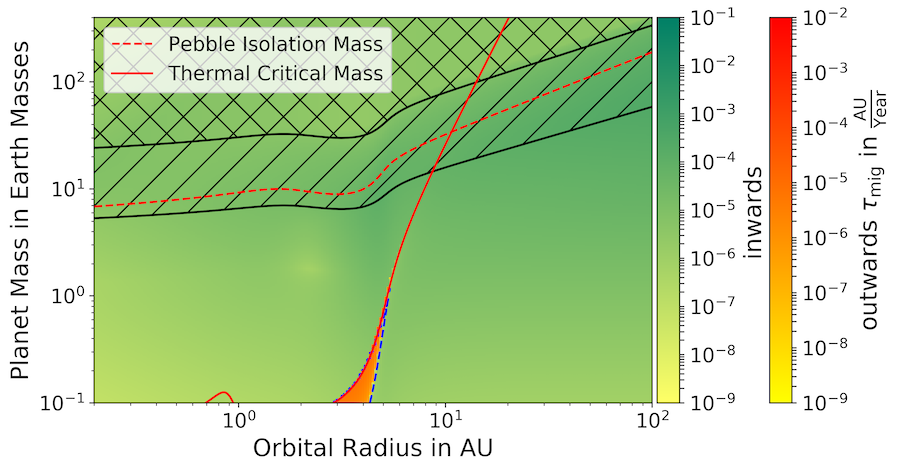}
                \end{subfigure}
                \subcaption{$\alpha = 10^{-3}$}
                \label{migrationmaps_visc-fig_visc1e-3-fig}
        \end{minipage}
        
        \begin{minipage}{\linewidth}
                \centering
                \begin{subfigure}{.44 \textwidth}
                        \centering
                        \includegraphics[width=\linewidth]{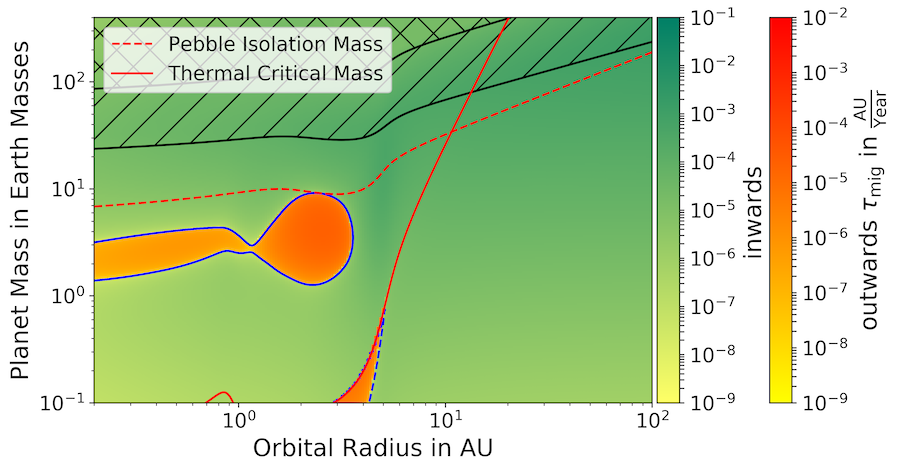}
                \end{subfigure}
                \begin{subfigure}{.44 \textwidth}
                        \centering
                        \includegraphics[width=\linewidth]{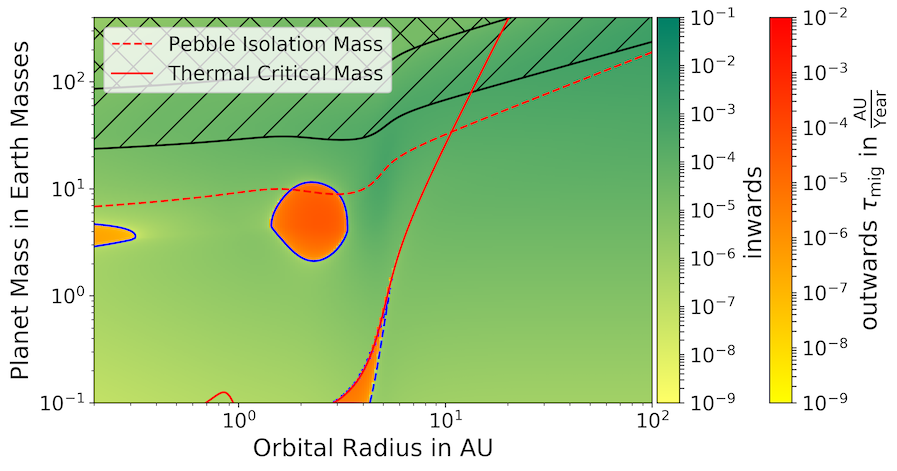}
                \end{subfigure}
                \subcaption{$\alpha = 5.4 \cdot 10^{-3}$}
                \label{migrationmaps_visc-fig_visc0-fig}
        \end{minipage}
        
        \begin{minipage}{\linewidth}
                \centering
                \begin{subfigure}{.44 \textwidth}
                        \centering
                        \includegraphics[width=\linewidth]{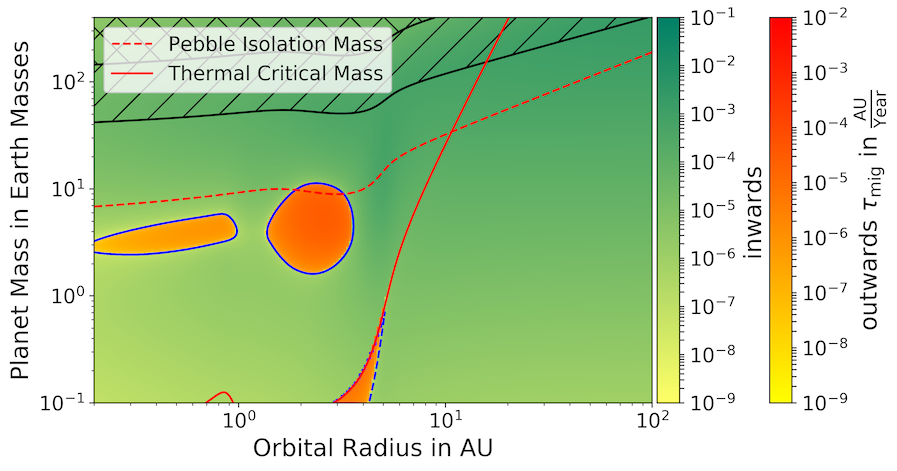}
                \end{subfigure}
                \begin{subfigure}{.44 \textwidth}
                        \centering
                        \includegraphics[width=\linewidth]{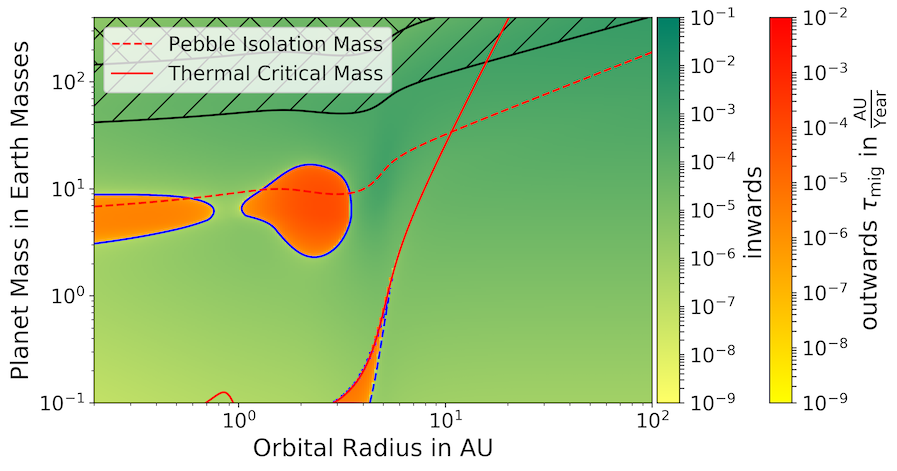}
                \end{subfigure}
                \subcaption{$\alpha = 10^{-2}$}
                \label{migrationmaps_visc-fig_visc1e-2-fig}
        \end{minipage}
        
        \caption{Migration maps after 1\,Myr with different $\alpha$-viscosity values for the calculation of migration timescales and the gap parameter.
                The left plots show the \cite{Paardekooper} formula and the right ones the \cite{ImprovedTypeI} formula. 
                Hatched regions mark the transition to type II migration, while planets in the cross-hatched areas are migrating completely in the type II regime. 
                Regions of outwards migration that are caused by the thermal torque, which is calculated with an accretion rate of \mbox{$\dot{M}_\mathrm{peb} \approx 1.4 \cdot 10^{-4} \, \frac{M_\mathrm{E}}{\mathrm{year}}$}, are outlined by dashed blue lines, while regions of outwards migration inside solid blue lines are due to the original torque formulae.
                Outwards migration in the original formulae is caused by the torques originating from the corotation region \citep{Paardekooper} or by the temperature torque \citep{ImprovedTypeI}.
                As type II migration depends on the disk properties rather the type I torque formula, the cross-hatched regions are identical between the corresponding left and right plots.
        }
        \label{migrationmaps_visc-fig}
        \vskip 0cm
\end{figure*}

The trends for changes in viscosity are applicable also to the \cite{Paardekooper} formula so we only show the results of the \cite{ImprovedTypeI} formula here.
Individual migration maps, however, show slightly different reactions to a change in viscosity.
This is shown in \autoref{migrationmaps_visc-fig}, where migration maps with both formulae at 1\,Myr are shown for different values of $\alpha$.
As mentioned before, the regions of outwards migration shrink with lower viscosity, until they disappear for $\alpha = 10^{-4}$, and the planets transition into type II migration at lower masses at low viscosity, since they can open gaps more easily.

The saturation functions of the corotation torque of the two models are based on the same physical requirements, but are slightly different numerically, similar to the torque components.
For a plot of the \cite{Paardekooper} saturation function see Figures 2 and 3 in \cite{Paardekooper}.

\section{Thermal torque from \cite{ThermalTorque}}
\label{thermal-section}
A recent development in type I migration is the thermal torque, which is a torque that acts on the planet due to its own heat release.
Two manifestations of this phenomenon have been discovered in the simulations of \cite{coldfinger} and \cite{2015Natur.520...63B}.

Thermal diffusion leads to two cold and dense lobes in the gas disk around the planet, which are asymmetric \citep{ThermalTorque}.
The resulting torque acting on the planet is called cooling torque or sometimes "cold finger effect" and was found in \cite{coldfinger}.
It is negative and accelerates inwards migration.

The other part of the thermal torque, which has been found in \cite{2015Natur.520...63B}, is the heating component.
Here, hot and underdense lobes form in the gas profile around luminous planets due to heat release.
This torque, again arising from asymmetry of the lobes, is positive and can lead to outwards migration.

The thermal torque acts on planets of up to intermediate mass only.
It gets cut off by two factors: When the planet reaches pebble isolation mass, solid accretion stops and the associated heat release vanishes.
In our model, we assume no planetesimal accretion which could still be ongoing after pebble isolation mass is reached.
However, the amount of impacting planetesimals at this stage is rather small (e.g.,\ \citealt{alibert2018formation}), so that not enough heat can be generated to fuel the heating torque.

The other cut-off mass is called thermal critical mass, which describes the point at which the planet stops generating excess internal energy outside the Bondi sphere, meaning the gas density is no longer perturbed.
This is estimated in \cite{ThermalTorque} as: 
\begin{equation}
M_\mathrm{crit,thermal} = \chi_\mathrm{P}c_{s,\mathrm{P}}/G,
\label{mCritThermal-formula}
\end{equation}
where $\chi_\mathrm{P}$ is the thermal diffusivity, $c_{s,\mathrm{P}} = h \cdot \Omega_\mathrm{P} \cdot r_\mathrm{P}$ is the local sound speed and $G$ is the gravitational constant.
The index P indicates that the quantities are evaluated at the planets' position

Our simulations use the description of the thermal torque developed by \cite{ThermalTorque}, which incorporates both the heating and cooling torque.
The luminosity for the heating torque is calculated as the solid accretion luminosity of the planet following \cite{Chrenko}.
This paper focuses on how the heating torque reacts to pebble accretion as the core growth model.
For details about calculations and formulae related to the thermal torque that are used here, see \autoref{thermal_torque_appendix}.

\subsection{Impact on migration}
The contribution of the thermal torque is already shown in \autoref{torque_comp-fig}, which shows the individual torque components and in \autoref{migrationmaps_visc-fig}, which shows the migration maps.
\autoref{torque_comp-fig} shows the typical course of the thermal torque (thick blue line).
As it is a composition of the negative cooling component and the positive heating part, it acts on planets with both signs.
Low-mass planets typically have a small accretion luminosity, so the cooling effect is the larger contribution.
Since the Hill radius scales with the third root of the core mass, planets can accrete enough pebbles for a positive thermal torque only when they get massive enough.
Finally, the thermal torque vanishes at thermal critical mass or at pebble isolation mass.
The thick purple lines show how the thermal torque changes the total torque according to the discussed behavior (the solid line includes the thermal torque, while the dashed one does not).

The migration maps in \autoref{migrationmaps_visc-fig} are only influenced by thermal torque below the two red lines which mark the cut-off.
Outwards migration due to the thermal torque is marked by dashed blue lines.
However, these new regions of outwards migration are small and restricted to low mass planets between 2 and 5\,AU.

The thermal torque is the same across all choices of $\alpha$, because the thermal component is independent of alpha, as we do not change the pebble scale height when we change alpha.
Instead, accretion rates were always calculated using $\alpha = 5.4 \cdot 10^{-3}$.
In a more realistic model, a lower viscosity reduces the pebble scale height, which results in larger accretion rates and thus a larger heating torque, potentially leading to larger regions of outwards migration due to the heating torque.
When we tested this dependence, however, we found that the impact on planet formation is still small.

\subsection{Impact on planet formation}
Similarly to the comparison between \cite{Paardekooper} and \cite{ImprovedTypeI}, migration maps show a significant change by the thermal torque, while the impact on planet formation is essentially negligible.

\autoref{migrationmaps_visc-fig} shows one of the main reasons why the thermal torque did not lead to major changes in planet formation in our model: the thermal critical mass cuts the thermal torque off at low masses in the inner disk.
This means that while outwards migration due to the heating torque does occur, most planets do not experience it during their formation.

However, this is only one part of the story, as the cut-off is much higher in the outer disk ($\approx 10 \mathrm{\ to\ } 100 \, M_\mathrm{E}$).
However, the thermal torque is inversely proportional to the thermal diffusivity, which typically increases when moving away from the star.
This means that while it is active, the thermal torque is small in the outer disk.
Additionally, pebble accretion is less efficient there, as the pebble surface density decreases and the pebble scale height increases with radius.
This means that the cooling torque outweighs the heating torque for planets that are far away from the star for the accretion rates used in this model, resulting in a low overall contribution of the heating torque to planet formation.
We discuss implications of different accretion rates in the next section.

\subsection{Thermal torque and pebble accretion}
As the heating torque is directly proportional to the solid accretion rate (see \autoref{accretion_luminosity-eq} and \autoref{heating_torque-eq}), the corresponding model for core growth plays a crucial role in determining the thermal torque.
The efficiency of pebble accretion depends on the amount of pebbles that are available to the planet.
This is regulated via the pebble flux $\dot{M}_\mathrm{peb}$ through the disk, which is typically set to \mbox{$\dot{M}_\mathrm{peb} = 2 \cdot 10^{-4} \cdot\exp (-\frac{t}{t_\mathrm{f}}) \, \frac{M_\mathrm{E}}{\mathrm{year}}$} (i.e.,\ 380\,$M_\mathrm{E}$ in total over 3\,Myr) in our model.
As discussed before, this is insufficient to yield a significant impact of the thermal torque on planet formation.

Research is typically directed towards slowing down migration to the star.
Therefore, it is interesting to look for positive contributions of the thermal torque.
\begin{figure*}[!htb]
        \centering
        \begin{minipage}{\linewidth}
                \begin{subfigure}{.49\textwidth}
                        \centering
                        \includegraphics[width=\linewidth]{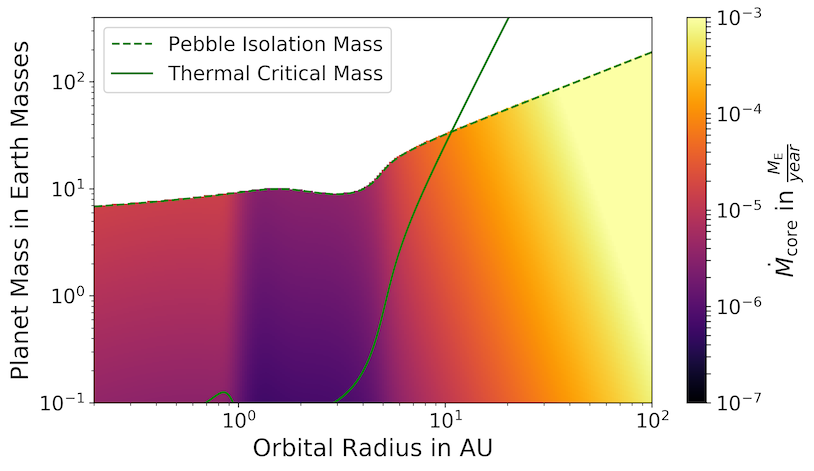}
                        Minimum core accretion rate
                \end{subfigure}
                \begin{subfigure}{.49\textwidth}
                        \centering
                        \includegraphics[width=\linewidth]{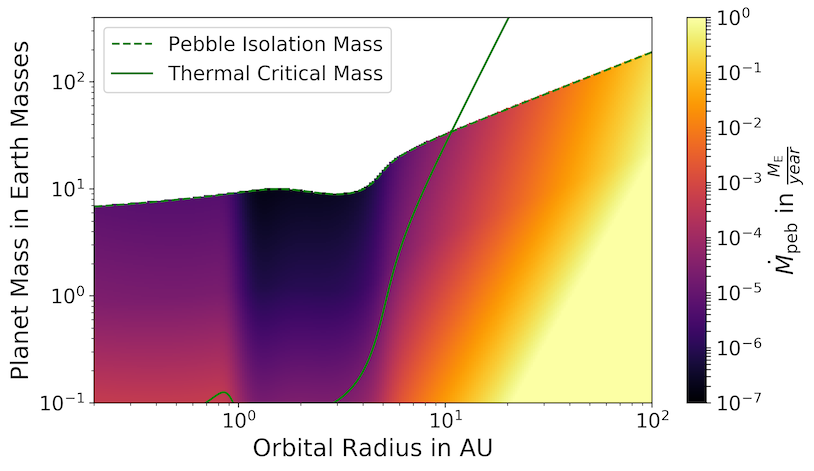}
                        Minimum pebble flux
                \end{subfigure}
        \end{minipage}
        
        \caption{Minimum core accretion rate and pebble flux that are required for a positive thermal torque in a disk that is 1\,Myr old. 
                The flux of pebbles needed to make a positive thermal torque also depends on the choice of Stokes number, which is set to 0.1 here.
                For a higher Stokes number, lower pebble fluxes are required for a positive thermal torque.
                The thermal critical mass used here is only an estimate, so values above might be of interest in the future.
                Below the thermal critical mass and in the outer disk, unreasonably high pebble fluxes are required for a positive thermal torque.
                Due to the exponential decrease in pebble flux that is employed in these simulations, about 72\% of the initial pebble flux are left after 1\,Myr, which corresponds to $1.4\cdot 10^{-4} \, \frac{M_\mathrm{E}}{\mathrm{year}}$ in the standard case.
                Keep in mind that a positive thermal torque does not automatically correspond to outwards migration.
                For different parameters of the disk age and how it influences the minimal pebble flux needed for a positive thermal torque, see  \autoref{mdotmin-fig_appendix} and \autoref{mdot_pebble_min-fig_appendix}.}
        \label{min_flux-fig}
\end{figure*}
\autoref{min_flux-fig} shows the minimum solid accretion rate and corresponding pebble fluxes that are required for a positive thermal torque at a given planetary mass and orbital distance using our disk model.
Keep in mind that this does not necessarily lead to outwards migration, as the Lindblad torque still needs to be overcome.
The plot extends beyond the thermal critical mass, as this is only an estimate in \cite{ThermalTorque} and regions of larger masses might become important in the future.

\begin{figure*}[!htb]
        \centering
        \begin{minipage}{\linewidth}
                \begin{subfigure}{.49\textwidth}
                        \centering
                        \includegraphics[width=\linewidth]{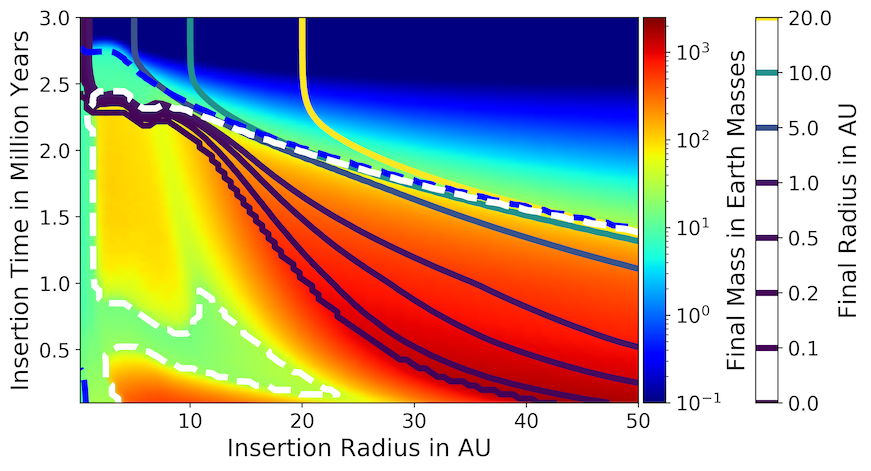}
                        Without thermal torque
                \end{subfigure}
                \begin{subfigure}{.49\textwidth}
                        \centering
                        \includegraphics[width=\linewidth]{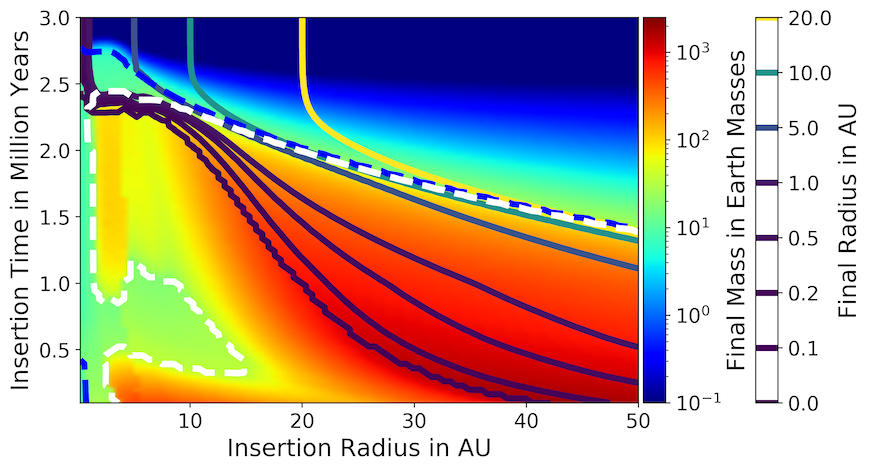}
                        Including thermal torque
                \end{subfigure}
        \end{minipage}
        
        \caption{Planet formation maps with and without the thermal torque from \cite{ThermalTorque} using the \cite{ImprovedTypeI} formula for the other torque components.
        Simulations were performed with an increased pebble flux of \mbox{$\dot{M}_\mathrm{peb} = 5 \cdot 10^{-4} \cdot\exp (-\frac{t}{t_\mathrm{f}}) \, \frac{M_\mathrm{E}}{\mathrm{year}}$, i.e.\ about 950\,$M_\mathrm{E}$ over 3\,Myr}.
        The plot should be read just like \autoref{planet_evo-fig}.
        Planets in the inner disk are significantly affected by the thermal torque due to the larger amount of pebbles.
        Planets that are inserted in the outer disk are not noticeably influenced by the thermal torque.}
        \label{increased_flux}
\end{figure*}

Clearly, \autoref{min_flux-fig} shows that unreasonably high pebble fluxes are required to achieve a meaningful positive contribution of the thermal torque.
However, it is still interesting to look at studies with increased amounts of pebbles, which can be observed in \autoref{increased_flux}.
Here, planet evolution maps are shown with a pebble flux of \mbox{$\dot{M}_\mathrm{peb} = 5 \cdot 10^{-4} \cdot\exp (-\frac{t}{t_\mathrm{f}}) \, \frac{M_\mathrm{E}}{\mathrm{year}}$} (i.e.,\ about 950 $M_\mathrm{E}$ in total, corresponding to a factor of 2.5 increase in pebbles to our nominal value), with and without the thermal torque.
While the contribution is still not large, it is noticeable at this pebble flux. 
Only planets that are inserted in the inner disk are affected, however.
This is due to the thermal torque being small in the outer disk and being cut-off for larger planets in the inner disk, meaning that planets that are inserted in the outer disk avoid regions of outwards migration caused by the heating torque during their growth.

When including the thermal torque, more planets reach runaway gas accretion.
This happens, for example, when they form at an early stage (e.g.,\ before 500\,kyrs).
This is caused by some outwards migration during the growth of the planets which has two effects: (i) as the planets move outwards, the pebble isolation mass increases so they can form larger cores which are more efficient at attracting gaseous envelopes; and (ii) as the planets are further away from the central star they need more time to migrate to the inner edge of the disk and thus have more time to grow.

\begin{figure*}[!htb]
        \centering
        \begin{minipage}{\linewidth}
                \begin{subfigure}{.49\textwidth}
                        \centering
                        \includegraphics[width=\linewidth]{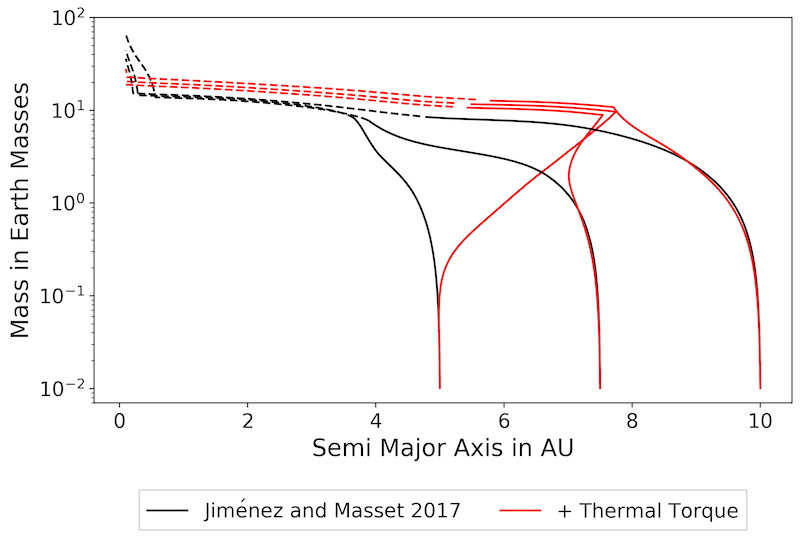}
                \end{subfigure}
                \begin{subfigure}{.49\textwidth}
                        \centering
                        \includegraphics[width=\linewidth]{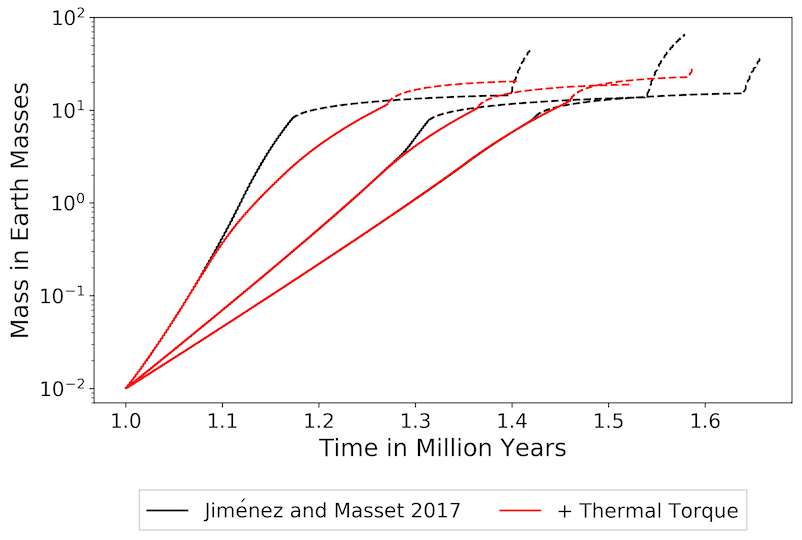}
                \end{subfigure}
        \end{minipage}
        
        \caption{Growth tracks of planets that have been inserted at different radii after 1\,Myr with the \cite{ImprovedTypeI} formula (no thermal torque) in black and the \cite{ImprovedTypeI} + \cite{ThermalTorque} formula (includes thermal torque) in red.
                The lines become dashed when the planets reach pebble isolation mass and are accreting gas.
                An increased pebble flux of $\dot{M}_\mathrm{peb} = 5 \cdot 10^{-4} \cdot \exp (-\frac{t}{t_\mathrm{f}}) \, \frac{M_\mathrm{E}}{\mathrm{year}}$ was used to achieve a larger contribution from the thermal torque.
                After performing significant outwards migration, planets that grow with the thermal torque migrate to the star quickly and grow less massive than the planets that were evolved without thermal torque, which reach runaway gas accretion.
                This might be due to the fact that heavier planets migrate faster, which gives them less time to contract an envelope before reaching the inner edge of the disk.}
        \label{heating_growth_tracks}
\end{figure*}

\autoref{heating_growth_tracks} shows select growth tracks with and without the thermal torque.
The planets have been inserted after 1\,Myr in a disk with a pebble flux of \mbox{$\dot{M}_\mathrm{peb} = 5 \cdot 10^{-4} \cdot\exp (-\frac{t}{t_\mathrm{f}}) \, \frac{M_\mathrm{E}}{\mathrm{year}}$} at various distances to the star.
Comparing it to \autoref{increased_flux}, we can see that this insertion scenario corresponds to planets that do not grow as much when including the thermal torque.
This is the case despite significant outwards migration and serves to show that planetary growth is subject to many variables.
Here, the planets spend less time in the regions of outwards migration created by corotation torques, since the initial outwards migration changes their trajectory in the migration maps and increases their pebble isolation mass, leading to fast inwards migration, where the planets have not enough time to contract their envelope and grow into gas giants before reaching the inner edge of the disk.

\section{Evaluation in the \cite{IdaDisk} disk model}
\label{idaDisk-section}
As discussed before, the disk model massively impacts planetary growth and migration.
This means that statements about migration in this paper are valid only in the disk models discussed here, rather than being generally applicable.
Therefore, it is useful to repeat the study for different disk environments.
Here, we discuss the implications of the heating torque when using the disk model from \cite{IdaDisk}.
Different components of the \cite{Paardekooper} torque formula and their influence on super-Earth formation has been discussed in \cite{brasser2018trapping}.

In order to adapt our code to the different disk, the temperature has been recalculated according to \cite{IdaDisk}, but all other variables remained the same.
In particular, stellar luminosity, viscosity and pebble accretion are modeled identically to the \cite{DiscModel} disk simulations.

Compared to the \cite{DiscModel} disk, this is a more simple model that employs power laws.
The temperature is seperated into two regimes: The inner disk, which is dominated by viscous heating and the outer disk, where stellar irradiation dominates the temperature profile.
As the disk evolves, the transition to stellar irradiation moves inwards.
The temperature is then calculated using:
\begin{gather} 
T_{\mathrm{vis}} \simeq 200 M_{* 0}^{3 / 10} \left( \alpha \cdot 10^3 \right) ^{-1 / 5} \left( \dot{M}_{*} \cdot 10^8 \right) ^{2 / 5}\left(\frac{r}{1\,\mathrm{AU}}\right)^{-9 / 10} \mathrm{K} ,\\ 
T_{\text {irr}} \simeq 150 L_{* 0}^{2 / 7} M_{* 0}^{-1 / 7}\left(\frac{r}{1 \text{\,AU}}\right)^{-3 / 7} \mathrm{K} ,\\
T=\max \left(T_{\mathrm{vis}}, T_{\mathrm{irr}}\right).
\end{gather}
We used $M_{*0} = 1.989\cdot 10^{33}$\,g for the stellar mass and the values from table 1 in \cite{DiscModel} for the luminosity of the star.
The gas surface density is then calculated through the gas accretion rate $\dot{M}$, which we assume to follow the same evolution as in the \cite{DiscModel} disk model.
The impact of the different disk model in terms of planet formation is quite significant, as we show in the following.

\subsection{Changes in migration maps}
Migration maps after 1\,Myr of disk evolution in the \cite{IdaDisk} disk model are shown in \autoref{Ida_migmaps}.
In the \cite{IdaDisk} disk, the \cite{ImprovedTypeI} formula produces larger regions of outwards migration than the \cite{Paardekooper} formula at the standard viscosity with $\alpha = 5.4\cdot 10^{-4}$.
This stands in contrast to the \cite{DiscModel} disk, where the opposite is true for most cases.

In fact, outwards migration is generally more scarce in the \cite{IdaDisk} disk than in the \cite{DiscModel} model.
This is particularly noticeable for the thermal torque, which yielded small regions of outwards migration in the \cite{DiscModel} disk, but only leads to essentially negligible amounts of outwards migration here.
The transition to type II migration is similar in the inner disk, where the temperature is set by viscous heating, but larger planetary masses are required in the outer disk.

\subsection{Changes in planet formation}
 \autoref{Ida_evomaps} presents planet evolution maps in the \cite{IdaDisk} disk model. Planets that evolve in the \cite{IdaDisk} model generally grow less massive than planets in the \cite{DiscModel} disk when considering the same pebble flux.
This is caused by the reduced region of outwards migration by the entropy related corotation torque.
In the \cite{DiscModel} model, the growing planets can be parked in this region of outwards migration and stay there until the end of the disk lifetime or they then slowly contract their envelope and grow to gas giants from this region.
Both options are not possible in the \cite{IdaDisk} model and planets instead migrate to the inner edge of the disk as super-Earths with a small gaseous envelope.
Apart from that, behavior is similar between the two models.

When comparing the \cite{Paardekooper} and the \cite{ImprovedTypeI} torque formulae, really similar planets are formed in these migration models and the thermal torque does not impact planet growth significantly at these pebble fluxes.
However, as the thermal torque produces even less outwards migration in this disk model, the impact on planet formation is also smaller when considering increased pebble fluxes.

\section{Discussion}
\label{discussion_section}
Here, we discuss the shortcomings and implications of our model and compare it with a previous study that incorporated the heating torque in planet formation simulations \citep{Guilera}.

\subsection{Comparison with \cite{Guilera}}
\label{guilera_section}
As we mentioned earlier in this paper, our results are only valid for the exact model that we tested.
This becomes clear when we compare our results to \cite{Guilera} , where the authors performed a similar study, but based on a different disk model.
Their results are very different, with the most important factor being that they found a very large level of impact on the part of the thermal torque.
In fact, they found very significant outwards migration of small planets due to the heating torque (see Figure 11 in \citealt{Guilera}), which strongly impacts their final mass.

Their model is different in two main ways:
As mentioned before, they operate in a different disk model, which goes beyond power laws for temperature (etc.)\ as well, but is very different from the \cite{DiscModel} model.
In their disk, the \cite{ImprovedTypeI} formula produces significantly less outwards migration, while migration maps employing the \cite{Paardekooper} formula look similar between the two disk models, that is, their work and ours.
The second major difference is that \cite{Guilera} employ a fixed core accretion rate of $10^{-5} \frac{M_\mathrm{E}}{yr}$ for the calculation of the heating torque.
We, on the other hand, use the solid accretion rate that follows from pebble accretion, which is usually significantly lower for small planets.

\cite{Guilera} come to the conclusion that the heating torque changes planetary growth and migration significantly.
We argue that they overestimated the impact of the heating torque dramatically when they employed the fixed core accretion rate of $10^{-5} \frac{M_\mathrm{E}}{yr}$ for planetary embryos.
This can be seen in \autoref{min_flux-fig}, where we show the minimal pebble flux and core accretion rate required to achieve a positive contribution of the thermal torque. 
In fact, in order to achieve the accretion rates used in \cite{Guilera}, we would need a pebble flux that is more than a factor of 10 higher than our highest used value. 
We deem this high pebble flux unrealistic for most stellar systems.

\subsection{Limitations of our model}
The planet growth model is very important for determining the planets' evolution.
As the heating torque is proportional to the solid accretion rate, the growth mechanism gains additional importance in our study of the thermal torque.
We only tested pebble accretion here, whereas a more realistic model would allow the planet to accrete planetesimals as well.
That said, for planetary embryos heavier than $10^{-2}\, M_\mathrm{E}$, pebble accretion is a lot more efficient than planetesimal accretion \citep{johansen2017forming}, so the inclusion of planetesimals is unlikely to change our results significantly.
\cite{johansen2019exploring} showed that planetesimal accretion can only be efficient in the inner disk, if planetesimals are small and stirring is low. 
This implies that planetesimal accretion can not be the main contribution to core accretion of giant planets at a few AU, as we investigate here.

Additionally, the fact that the solid accretion rate depends on the $\alpha$-viscosity via the pebble scale height (see \autoref{pebble_scale_height} and \autoref{3D_pebble_accretion-eq}) was neglected.
At lower $\alpha,$ the pebble scale height decreases and planets enter the more efficient 2D-pebble accretion regime more quickly.
Additionally, the reduction factor in the three dimensional accretion rates scales with $\alpha^{-1/2}$. 
That means at the lowest value that we tested ($\alpha=10^{-4}$), solid accretion rates in the 3D-regime are larger by a factor of 10 compared to the highest value that we tested ($\alpha=10^{-2}$).
However, when we implemented this dependence we found that while planets grow significantly more massive and outwards migration from the heating torque was stronger, the influence of the thermal torque on planet evolution was still small.

The main reason why the thermal torque did not change migration much in our simulations was that the thermal critical mass cuts it off at low masses in the inner disk.
It is important to note that this thermal critical mass was only an estimate of when the planet stops generating excess internal energy outside the Bondi sphere \citep{ThermalTorque}.
Should future research show that the thermal critical mass is in fact larger than we assumed here, our study of the thermal torque would have to be repeated.

\subsection{Influence in N-body simulations}
A more impactful simplification is the evolution of single planets only. 
As mentioned before, planet-planet interactions have a huge impact on planetary migration and are not captured here.

In particular, even though the heating torque has only minimal consequences on the overall formation of planets in our systems, the migration directions and speeds can change significantly (\autoref{heating_growth_tracks}). 
This has important consequences for the trapping of planets in resonances, because the relative migration speed between the planets determines in which resonance the planets will be trapped.
It is extremely difficult to asses the general implications of the heating torque for N-body simulations with our current simulations. 
In order to determine the influence of the torque formulae and the thermal torque on super-Earth systems, where the trapping in resonances plays an important role (e.g., \citealt{outwardImportance2, Izidoro2019, lambrechts2019formation}), N-body simulations featuring these different prescriptions have to be undertaken.

\subsection{Influence on the chemical composition of planets}
When a planet crosses an ice line in the disk, the chemical composition of the accreted material changes (e.g., \citealt{oberg2011effects, madhusudhan2017atmospheric}). 
In particular, \cite{bitsch2019rocky} showed that the direction of migration close to the ice lines has important consequences for the chemical composition of formed planets.

The heating torque has the potential to change the direction of migration if the accretion rate onto the planet is large. 
This implies that planets forming around metal-rich stars could have different compositions due to the larger influence of the heating torque compared to planets forming around metal poor stars. 
Thus, we think that the investigation of the heating torque in high metallicity environments could be crucial if planetary compositions can be linked to the planet formation pathways.

\section{Conclusion}
\label{conclusion_section}
Here, we present a brief summary of our results.
When comparing the type I torque formulae from \cite{Paardekooper} and \cite{ImprovedTypeI}, we found that they produce very similar planets in a single planet formation scenario (\autoref{planet_evo-fig}), despite significant differences in the migration maps (\autoref{migrationmaps_visc-fig}).

Adding the thermal torque following \cite{ThermalTorque} introduces new regions of outwards migration to the migration maps due to the heating component (\autoref{migrationmaps_visc-fig}).
The addition of the thermal torque to either of the two type I torque formulae investigated here does not yield a significant impact on single planet formation (\autoref{torque_comp-fig} and \autoref{increased_flux}), however, as outwards migration caused by the heating torque is limited to very small planets and a small interval of distances to the star around 3\,AU (\autoref{migrationmaps_visc-fig}).

The specific influence on the planetary migration rates with the thermal torque crucially depends on the underlying disk model and on the accretion rate, where we followed an accretion recipe based on pebble accretion (e.g., \citealt{lambrechts2014forming, PebbleAccretion}).
 If the accretion rates onto the planet are larger, outwards migration can be achieved, even though it is difficult to achieve these accretion rates with realistic dust-to-gas ratios (\autoref{min_flux-fig}).
 However, even then the influence on planet formation seems minimal in our simulations (\autoref{increased_flux}).
 Summing up, the main conclusion of this paper is that thermal torque only has a minimal impact on planetary growth.

The core growth model that we used here is pebble accretion.
Other models will result in different accretion rates, which changes not only the trajectory in the migration maps and, hence, planet evolution in general, but also the heating torque.

While the difference between the type I torque formulae is small, it is still advisable to employ the \cite{ImprovedTypeI} \mbox{type I} torque formula as it is derived in a three-dimensional disk, which is a more realistic approach than the two-dimensional simulations from \cite{Paardekooper}.
Hence, the \cite{ImprovedTypeI} formula should produce more realistic migration rates than the \cite{Paardekooper} formula.

The heating torque starts affecting planet formation at high pebble fluxes so it should be included in simulations of planets forming around high metallicity stars.
Planets formed around low-metallicity stars, on the other hand, will barely exhibit any impact from the heating torque.

\begin{acknowledgements}
        T.Baumann and B.Bitsch thank the European Research Council (ERC Starting Grant 757448-PAMDORA) for their financial support and the referee for their helpful comments that improved the manuscript.
\end{acknowledgements}

%
\bibliographystyle{aa} 
\bibliography{bibliography} 
%

\begin{appendix}
\section{Type I migration model from \cite{Paardekooper}}
\label{Paardekooper_torque-appendix}
The torque formulae of \cite{Paardekooper} are the result of an analysis of 2D hydrodynamical simulations in non-barotropic disks with finite viscosity and constant thermal diffusion.
Furthermore, they use simple power laws in radius for surface density $\Sigma$, temperature $T$ and viscosity $\nu$.
They set the background temperature profile as an equilibrium of viscous heating and radiative cooling. 
They found that the formulae they derived agreed with their simulations up to a maximum error of 20\%.

All torque components are normalized to $\Gamma_0$ (\autoref{Gamma_0-eq}).
We note that $\Gamma_0$ depends quadratically on the planet mass via $q$, so small planets migrate more slowly than more massive ones.
As a result, the main part of the migration happens just before the planet becomes massive enough to open a gap which slows the migration down again as the planet enters type II migration.
They also used an effective adiabatic index $\gamma_\mathrm{eff}$ which modifies the adiabatic index according to the planets position, the disk aspect ratio and thermal diffusivity according to
\begin{equation}
\label{gamma_eff-eq}
\gamma_\mathrm{eff} = \frac{2 Q \gamma}{\gamma Q+\frac{1}{2} \sqrt{2 \sqrt{\left(\gamma^{2} Q^{2}+1\right)^{2}-16 Q^{2}(\gamma-1)}+2 \gamma^{2} Q^{2}-2}},
\end{equation}
where,
\begin{equation}
Q=\frac{2 \chi_{\mathrm{p}}}{3 h^{3} r_{\mathrm{p}}^{2} \Omega_{\mathrm{p}}},
\end{equation}
where $\chi$ is the thermal diffusivity.
This effective adiabatic index aims to account for regions in the disk that can cool efficiently, that is,\ the outer disk, producing isothermal results, while regions that can not cool efficiently in the inner disk produce adiabatic results.
The width of the horseshoe region is determined as: 
\begin{equation}
\label{paardekooper_xs_formula}
x_s = \frac{1.11}{\gamma_\mathrm{eff}^{1/4}} \sqrt{\frac{q}{h}} \cdot r_\mathrm{P}.
\end{equation}

In this model, the saturation of the corotation torque is tackled via a modification of \cite{Masset_2001}.
When the viscosity gets very large, the viscous diffusion timescale is no longer much shorter than it takes for material that has performed a horseshoe turn to enter back into the horseshoe region.
In this case, the horseshoe drag gets replaced by a linear corotation torque, which is typically smaller.
Since this scenario requires high viscosity, saturation no longer occurs and the torque becomes fully linear.
The saturation is modeled via a function $F$ in \autoref{corotation_torque-formula}, which can be seen in Figures 2 and 3 in \cite{Paardekooper}.
Generally, as the viscosity increases, the horseshoe drag becomes smaller and the linear corotation torque gets larger.
These two processes take place independently on different timescales and they are modeled by two different functions $G \geq 0$ and $K \leq 1$ such that:
\begin{equation}
\label{corotation_torque-formula}
\Gamma_\mathrm{c} = G(p) \cdot F(p) \cdot \Gamma_\mathrm{hs} + (1-K(p)) \cdot \Gamma_\mathrm{c,lin},
\end{equation}
where $p$ is a saturation parameter that depends on the planets' position, its horseshoe width, and the viscosity.
$G$ and $K$ follow the same general formula, but with different time scale parameters such that towards low viscosities the linear torque decreases slower than the horseshoe drag increases.

The saturation function $F$ for the horseshoe drag is:
\begin{equation}
\label{Paardekooper_saturation_eq}
F(p)=\frac{8 I_{4 / 3}(p)}{3 p I_{1 / 3}(p)+\frac{9}{2} p^{2} I_{4 / 3}(p)}
,\end{equation}
where $I_{4 / 3}$ is a modified Bessel function.
The parameter that is entered into this function is:
\begin{gather}
p=2 \sqrt{k x_{\mathrm{s}}^{3}} / 3
\\ k=\frac{r_{\mathrm{p}}^{2} \Omega_{\mathrm{p}}}{2 \pi \nu_{\mathrm{p}}}.
\end{gather}
The entropy torque saturates not only with viscosity but also with thermal diffusion, in this case the saturation parameter is:
\begin{gather}
p_\chi=2 \sqrt{k_\chi x_{\mathrm{s}}^{3}} / 3
\\ k_\chi=\frac{r_{\mathrm{p}}^{2} \Omega_{\mathrm{p}}}{2 \pi \chi_{\mathrm{p}}}.
\end{gather}

The blending functions $G$ and $K$ follow
\begin{gather}
G(p)=\left\{\begin{array}{cc}{\frac{16}{25}\left(\frac{45 \pi}{8}\right)^{3 / 4} p^{3 / 2}} & {p<\sqrt{\frac{8}{45 \pi}}} \\ {1-\frac{9}{25}\left(\frac{8}{45 \pi}\right)^{4 / 3} p^{-8 / 3}} & {p \geq \sqrt{\frac{8}{45 \pi}}}\end{array}\right.
\\ K(p)=\left\{\begin{array}{cc}{\frac{16}{25}\left(\frac{45 \pi}{28}\right)^{3 / 4} p^{3 / 2}} & {p<\sqrt{\frac{28}{45 \pi}}} \\ {1-\frac{9}{25}\left(\frac{28}{45 \pi}\right)^{4 / 3} p^{-8 / 3}} & {p \geq \sqrt{\frac{28}{45 \pi}}}\end{array}\right.
\end{gather}

\subsection*{Lindblad torque}
\label{lindblad_paardekooper-section}
The formula for the Lindblad torque found by \cite{Paardekooper} is
\begin{equation}
\label{Paardekooper_Lindblad-formula}
\gamma_\mathrm{eff} \frac{\Gamma_\mathrm{L}}{\Gamma_0} = -2.5 - 1.7\beta + 0.1\alpha
,\end{equation}
where $\gamma_\mathrm{eff}$ is the effective adiabatic index from \autoref{gamma_eff-eq}, $-\beta$ is the temperature exponent and $-\alpha$ is the surface density exponent.
Usually, both temperature and surface density decrease with the distance to the star so $\alpha$ and $\beta$ are typically positive.
For the disk model employed here, which is the one from \cite{DiscModel}, that is not always the case, however.
The Lindblad torque is typically the main negative contribution and drives fast inward migration, as we can see in \autoref{torque_comp-fig}.

\subsection*{Barotropic torque}
The barotropic torque is the result of the horseshoe trajectories which the material in the corotation region follows.
The planet pushes material forwards that comes from the outer disk and pushes material backwards that comes from the inner disk.
As a result of the surface density gradient in the disk, the planet typically pushes more mass backwards and migrates outwards as a consequence.
Naturally, this torque component depends on the surface density gradient.
The unsaturated barotropic torque from the horseshoe drag follows 
\begin{equation}
\gamma_\mathrm{eff} \frac{\Gamma_\mathrm{hs,baro}}{\Gamma_0} = 1.1  \bigg ( \frac{3}{2} - \alpha \bigg )
\end{equation}
and the linear barotropic corotation torque is given by
\begin{equation}
\gamma_\mathrm{eff} \frac{\Gamma_\mathrm{c,lin,baro}}{\Gamma_0} = 0.7 \bigg ( \frac{3}{2} - \alpha \bigg ).
\end{equation}
This part of the torque does not depend on thermal diffusion.
As the surface density generally drops with increasing radius, $\alpha$ is positive, and the barotropic torque gets positive when the gradient is shallow.

\subsection*{Entropy torque}
In this torque model the entropy torque is the main component responsible for outwards migration.
It relies on the horseshoe trajectories of the corotating material like the barotropic torque, but is a consequence of the temperature difference between the inner and outer disk.
Material that gets pushed from the outer disk to the inner disk typically gets warmer and contracts, leading to a greater gravitational pull in front of the planet than the material behind the planet, which cools down and expands.
The entropy related horseshoe drag is:
\begin{equation}
\label{pa_entropy_hs-eq}
\gamma_\mathrm{eff} \frac{\Gamma_\mathrm{hs,ent}}{\Gamma_0} = 7.9 \frac{\xi}{\gamma}
\end{equation}
and the linear entropy corotation torque is
\begin{equation}
\label{pa_entropy_lin-eq}
\gamma_\mathrm{eff} \frac{\Gamma_\mathrm{c,lin,ent}}{\Gamma_0} = \bigg ( 2.2  - \frac{1.4}{\gamma} \bigg ) \xi
,\end{equation}
\nopagebreak
with $\xi = \beta - (\gamma - 1) \alpha$ as the entropy gradient.
Here, we note that $\gamma$ is used, and not $\gamma_\mathrm{eff}$, because the temperature gradient of the disk is unrelated to the cooling properties of the planet.
The entropy torque needs both thermal and viscous diffusion to remain unsaturated.
Thermal diffusion is required to restore the entropy gradient and viscous diffusion is required to diffuse the vortensity since the entropy torque originates from a streamline near the seperatrix of the horseshoe region.
Consequently, \autoref{corotation_torque-formula} needs to be modified for the entropy torque as $F(p) \rightarrow F(p_\nu)F(p_\chi)$, with viscosity, $\nu,$   and thermal diffusivity, $\chi$  .Values for 
$p_\nu$ and $p_\chi$ depend on the planet's position, horseshoe width, and on the viscosity or thermal diffusivity, respectively.
Similarly, the transition functions $G$ and $K$ in \autoref{corotation_torque-formula} only get modified by choosing new arguments: $G(p) \rightarrow \sqrt{G(p_\nu)G(p_\chi)}$ and $1-K(p) \rightarrow \sqrt{(1-K(p_\nu))(1-K(p_\chi))}$.
Since a higher thermal diffusivity keeps advection of entropy at the separatrix at bay, it increases the torque initially.
However, when $\chi$ becomes too large, the disk essentially behaves locally isothermal where the corotation torque is given by the barotropic horseshoe drag and the linear part of the entropy torque.
As the non-linear part of the entropy component is larger than the linear one, this leads to a smaller overall torque.

\section{Type I migration model from \cite{ImprovedTypeI}}
\label{JM_torque-appendix}
\cite{ImprovedTypeI} claim that the formulae from \cite{Paardekooper} are not accurate enough for planet population synthesis. 
They argue that although the key mechanisms have been captured, many constants in the formulae are off.
The main update compared to \cite{Paardekooper} is that these considerations are based on three-dimensional hydrodynamical simulations rather than just two-dimensional ones.
They also update the horseshoe width to remain valid for more intermediate mass planets of several Earth-masses.
The new formula for the horseshoe width $x_s$ is:
\begin{equation}
\label{jm_xs_formula}
x_s = \frac{1.05(q/h)^{\frac{1}{2}} + 3.4  q^{\frac{7}{3}}/h^6}{1+2 q^2/h^6} r_\mathrm{P}.
\end{equation}
They do, however, still find some discrepancies between their formulae and simulations.
They base their reasoning on multiple simplifications, including the neglect of the influence of the planet's torque on the disk.
Before the planet opens a full gap, it starts perturbing the density profile around it, which can alter the effect of the Lindblad torque and can change the thermal diffusivity, but this has not been taken into consideration here.

Rather than simplifying the formulae as far as possible, \cite{ImprovedTypeI} kept one torque component for each source of torque.
This results in four corotation components ($\Gamma_\mathrm{C} = \Gamma_\mathrm{V} + \Gamma_\mathrm{S} + \Gamma_\mathrm{T} + \Gamma_\mathrm{VCT})$ and one Lindblad component $\Gamma_\mathrm{L}$, all displayed in \autoref{torque_comp-fig}.
Comparing the thick dashed purple lines in \autoref{torque_comp-fig}, which mark the total torque acting on the planet (without thermal torque), we can see that both follow similar curves, but are different in terms of the masses at which they act and in their magnitude.
This can have a significant influence on migration. 

Most torque components are still normalized to $\Gamma_0$ from \autoref{Gamma_0-eq}.
Also, as before, the corotation torque components are divided into two subcomponents, the linear one and the horseshoe drag which can saturate.
Rather than developing a new saturation mechanism, they use the one from \cite{Masset_2010}.
A notable difference to \cite{Paardekooper} is that the blending of the subcomponents happens at the same pace, whereas \cite{Paardekooper} assume the linear part to decrease slower than the horseshoe part increases as the viscosity decreases.

\subsection*{Lindblad torque}
\label{lindblad_JM-section}
The Lindblad torque was taken from \cite{Tanaka_2002} and reexamined in a locally isothermal setup.
As it was only the coefficient for the temperature dependence that was reworked, the idea was to test the torque on a planet that was in the disk long enough for the corotation torque to saturate, leaving only the Lindblad torque and to fit the result.
The new formula is: 
\begin{equation}
\label{JM_Lindblad-formula}
\frac{\Gamma_\mathrm{L}}{\Gamma_0} = -(2.34 - 0.1\alpha+1.5\beta) \cdot f(\frac{\chi}{\chi_\mathrm{c}})
,\end{equation}
with
\begin{equation}
\label{JMf-formula}
f(x) = \frac{\sqrt{\frac{x}{2}} + \frac{1}{\gamma}}{\sqrt{\frac{x}{2}}+1}.
\end{equation}
The factors 2.34 and -0.1 in \autoref{JM_Lindblad-formula} were taken from \cite{Tanaka_2002}, $\chi_\mathrm{c} = r^2_\mathrm{P}h^2\Omega_\mathrm{P}$ is the critical thermal diffusivity from \cite{Masset_2010}.
The function $f(\frac{\chi}{\chi_\mathrm{c}})$ in \autoref{JMf-formula} acts like the effective adiabatic index $\gamma_\mathrm{eff}$ in \autoref{Paardekooper_Lindblad-formula}.

\subsection*{Entropy torque}
Like in \cite{Paardekooper}, the entropy torque is the result of an entropy gradient in the disk.
However, this part has been updated in a meaningful way.
Similar to \cite{Paardekooper}, the entropy torque is mixed between the linear and horseshoe part by both the viscosity and the thermal diffusivity so there is an $\epsilon_\nu$ and an $\epsilon_\chi$ to determine the blending.
The formula is:
\begin{equation}
\label{jm_entropy-eq}
\Gamma_\mathrm{S} = \epsilon_\nu \epsilon_\chi \Gamma_\mathrm{S}^\mathrm{hs} + (1-\epsilon_\nu \epsilon_\chi)\Gamma_\mathrm{S}^\mathrm{lin}.
\end{equation}
The blending coefficients are taken from \cite{Masset_2010} and amount to:
\begin{gather}
\label{epsilon_nu-formula}
\epsilon_\nu = \left( 1+\left(6 h z_{\nu}\right)^{2} \right)^{-1}
,\\ \epsilon_\chi = (1 + 15hz_\chi)^{-1}
,\\ z_\chi       = \frac{r_\mathrm{P}\chi}{\Omega_\mathrm{P}x_s^3}
,\\ z_\nu = \frac{r_\mathrm{P} \nu}{\Omega_\mathrm{P} x_s^3}.
\label{z_nu-eq}
\end{gather}
The linear and unsaturated horseshoe parts follow:
\begin{gather}
\gamma \frac{\Gamma_\mathrm{S}^\mathrm{lin}}{\Gamma_0} = 0.8 \xi'
,\\ \Gamma_\mathrm{S}^\mathrm{uhs} = 3.3 \xi' \Sigma_0 \Omega_\mathrm{P} x_s^4
,\\ \xi' = \beta - 0.4 \alpha - 0.64.
\label{xi-formula}
\end{gather}
We note that  $\Gamma_\mathrm{S}^\mathrm{uhs}$, as in the following unsaturated horseshoe torques, is not given in terms of $\Gamma_0$.
The quantity $\xi'$ is called entropy gradient by \cite{ImprovedTypeI} because it shows resemblance to the actual radial entropy gradient.
The horseshoe drag is the product of the unsaturated formula and a saturation function $F_\mathrm{S}$.
\begin{gather}
\Gamma_\mathrm{S}^\mathrm{hs} = F_\mathrm{S} \Gamma_\mathrm{S}^\mathrm{uhs}
,\\ F_\mathrm{S} = 1.2 \cdot \overline{1.4z_\chi^{1/2}} \cdot \overline{1.8z_\nu^{1/2}},
\end{gather}
where $\overline{x} = \mathrm{min}(1, x)$.

\subsection*{Temperature torque}
This torque component is a result of the temperature gradient in the midplane of the disk and was incorporated into a type I torque recipe for the first time in \cite{ImprovedTypeI}.
It is found in locally isothermal simulations with vanishing entropy and surface density gradients but $\beta \neq 0$.
The temperature gradient produces vortensity in isothermal calculations which is concentrated at the downstream seperatrices.
It is reasonable to assume that the thus generated torque depends on viscosity in the same way as the entropy torque.
The formula is:
\begin{equation}
\Gamma_\mathrm{T} = \epsilon_\nu \Gamma_\mathrm{T}^\mathrm{hs} + (1-\epsilon_\nu)\Gamma_\mathrm{T}^\mathrm{lin}
,\end{equation}
with the mixing coefficient $\epsilon_\nu$ from \autoref{epsilon_nu-formula}.
Once again, the temperature component uses a different saturation function and linear component:
\begin{gather}
\gamma \frac{\Gamma_\mathrm{T}^\mathrm{lin}}{\Gamma_0} = 1.0 \beta
,\\ \Gamma_\mathrm{T}^\mathrm{hs} = F_\mathrm{T}(z_\nu) \Gamma_\mathrm{T}^\mathrm{uhs}
,\\ \Gamma_\mathrm{T}^\mathrm{uhs} = 0.73 \beta \Sigma_0 \Omega_\mathrm{P} x_s^4
,\\ F_\mathrm{T} = 1.2 \cdot \overline{1.8 z_\nu^{1/2}}.
\end{gather}

\subsection*{Vortensity torque}
The vortensity torque was not updated by \cite{ImprovedTypeI} since the horseshoe drag has the same expression in two and three dimensions in a barotropic disk.
This component corresponds to the barotropic torque in \cite{Paardekooper}.
To simplify, they assume a vortensity gradient of $\frac{3}{2} - \alpha$, which is only valid for disks in which the surface density and angular frequency are power laws with respect to the distance to the star.
The specific formula is: 
\begin{equation}
\Gamma_\mathrm{V} = \epsilon_\mathrm{b}\Gamma_\mathrm{V}^\mathrm{hs} + (1-\epsilon_\mathrm{b})\Gamma_\mathrm{V}^\mathrm{lin}.
\end{equation}
With the blending coefficient:
\begin{align}
\epsilon_\mathrm{b} = (1 + 30 h z_\nu)^{-1}
\end{align}
from \cite{Masset_2010}.
The linear component is:
\begin{equation}
\gamma \frac{\Gamma_\mathrm{V}^\mathrm{lin}}{\Gamma_0} = 0.976 - 0.640 \alpha. 
\end{equation}
The horseshoe drag component is given by: 
\begin{equation}
\Gamma_\mathrm{V}^\mathrm{hs} = \frac{8\pi}{3}z_\nu F(z_\nu) \Gamma_\mathrm{V}^\mathrm{uhs}
,\end{equation}
where $\Gamma_\mathrm{V}^\mathrm{uhs}$ is the unsaturated horseshoe drag and $F(z_\nu)$ in combination with the prefactors is the saturation function.
The remaining quantities are:
\begin{equation}
\Gamma_\mathrm{V}^\mathrm{uhs} = \frac{3}{4} \bigg ( \frac{3}{2} - \alpha \bigg ) \Sigma_0 \Omega_\mathrm{P}^2 x_s^4
,\end{equation}
\[
F(x)= 
\begin{cases}
1 - x^\frac{1}{2},& \text{if } x < \frac{4}{9}\\
\frac{4}{27x},              & \text{otherwise.}
\end{cases}
\]

\subsection*{Viscous coupling torque (VCT)}
This part results from viscous production of vortensity at the density jumps at the seperatrices of the horseshoe region.
As this phenomenon largely depends on the entropy gradient it is assumed to scale with the quantity $\xi'$ from \autoref{xi-formula}, which takes the role of an entropy gradient.
However, they normalize $\xi'$ by $\beta=1$ for this term.
Unlike the other corotation torques, this component does not have a linear equivalent.
\cite{Masset_2010} have proposed that $\Gamma_\mathrm{VCT}$ decays like the vortensity torque with the mixing coefficient $\epsilon_\mathrm{b}$.
The formula is:
\begin{equation}
\Gamma_\mathrm{VCT} = \frac{4 \pi \xi}{\gamma} \Sigma_0 \Omega_\mathrm{P} x_s^4 \epsilon_\mathrm{b} z_\nu \frac{z_\nu F(z_\nu) - z_\chi F(z_\chi)}{z_\nu - z_\chi}.
\end{equation}
In our the simulations, the viscous coupling term is found to have an almost negligible influence on the total torque.

\section{Thermal torque from \cite{ThermalTorque}}
\label{thermal_torque_appendix}
Here, the thermal torque consists of two components: a cooling torque which is negative and a heating torque which is positive.
Both act as a result of density perturbations in the gas profile around the planet due to heat exchange with the disk.
In this paper, the planets luminosity is its accretion luminosity caused by the pebble flux onto the planet.
The typical course of the thermal torque is visible in \autoref{torque_comp-fig} (thick blue line).

\begin{figure}[!thb]
        \centering
        \begin{subfigure}{0.49\textwidth}
                \centering
                \includegraphics[width=\linewidth]{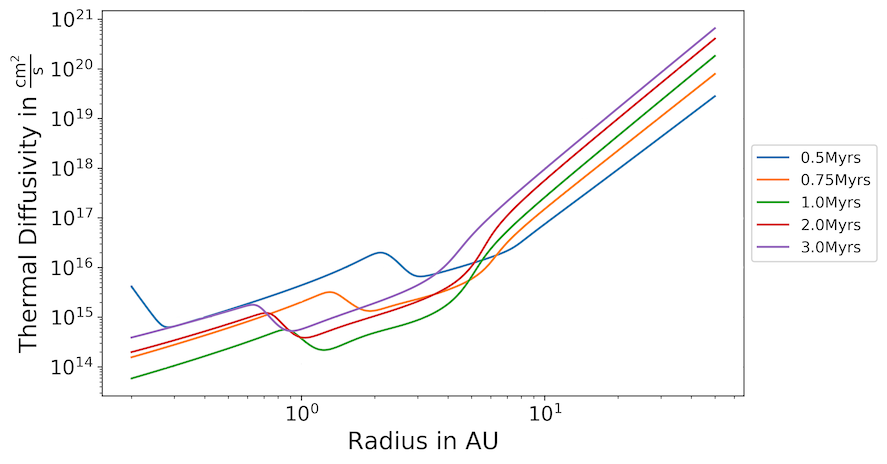}
                \caption{Thermal diffusivity}
                \label{thermal_diffusivity-fig}
        \end{subfigure}
        \begin{subfigure}{.49\textwidth}
                \centering
                \includegraphics[width=\linewidth]{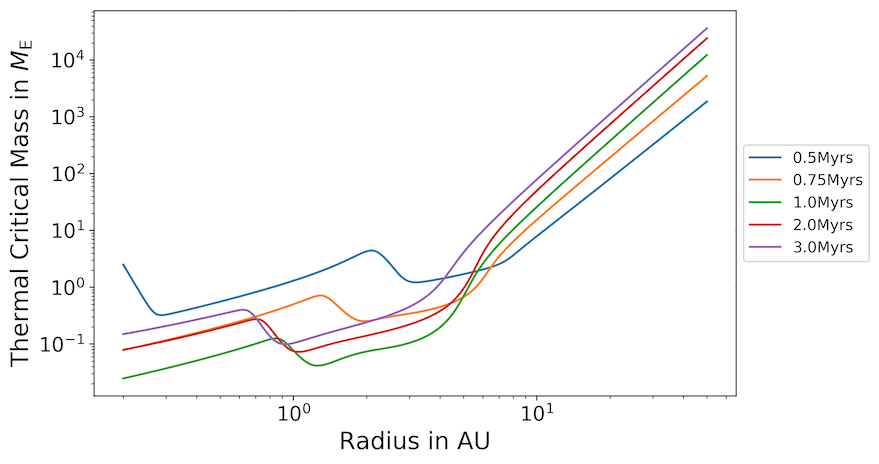}
                \caption{Thermal critical mass}
                \label{thermal_critical_mass-fig}
        \end{subfigure}
        \caption{Disk thermal diffusivity and thermal critical mass over distance to the star at various times with a pebble flux of $2 \cdot 10^{-4} \cdot\exp (-\frac{t}{t_\mathrm{f}}) \, \frac{M_\mathrm{E}}{\mathrm{yr}}$. The change in thermal diffusivity dominates the change in thermal critical mass. The other changing variable in the thermal critical mass is the aspect ratio.
        }
        \label{thermal_critical-fig}
\end{figure}

One source of the cut-off effect on the thermal torque, the thermal critical mass, is plotted in \autoref{thermal_critical_mass-fig} at different times in the disk from \cite{DiscModel}.
The evolution over time is dominated by the thermal diffusivity, which changes by several orders of magnitude (\autoref{thermal_diffusivity-fig}).
The disk aspect ratio also changes over time, but the influence is smaller.
It should be noted that the thermal critical mass (\autoref{mCritThermal-formula}) is only an estimate in \cite{ThermalTorque} and needs to be investigated in more detail in the future, especially as it becomes very small at later times of disk evolution, which has important consequences for the thermal torque.

\begin{figure}[!thb]
        \centering
\includegraphics[width=\linewidth]{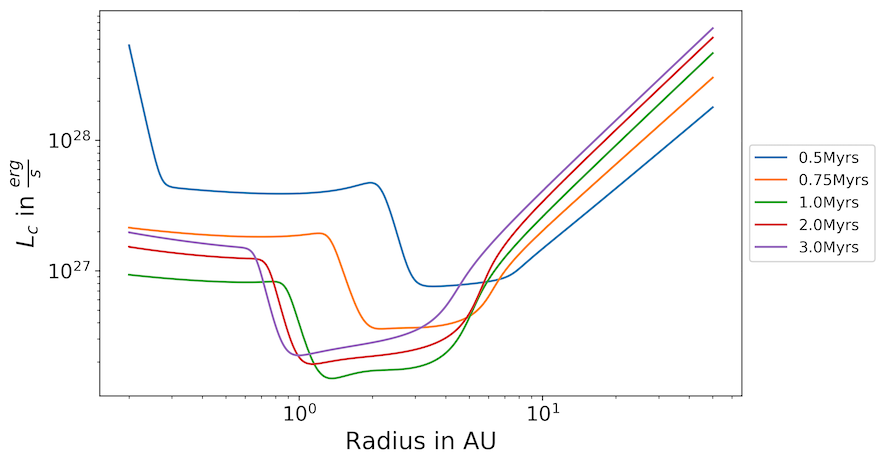}
\caption{Critical luminosity}
\label{lc_thermal-fig}
\caption{Critical luminosity (\autoref{L_c-eq}) for a planet mass of 1\,$M_\mathrm{E}$ (right) at various times. The kinks in the critical luminosity are caused by the thermal diffusivity (see \autoref{thermal_diffusivity-fig}).
}
\label{L_crit-fig}
\end{figure}

An important ingredient for both components is a finite thermal diffusivity and the simulations can not be adiabatic as well.
\cite{ThermalTorque} derive their formulae from investigating low mass planets in a disk with thermal diffusivity with linear perturbation theory in a three-dimensional shearing sheet.
Their results for the cooling torque are:
\begin{gather}
\gamma\frac{\Gamma_\mathrm{thermal}^\mathrm{cold}}{\Gamma_0} = -1.61 (\gamma -1) \frac{x_p}{\lambda_c}
\\ \lambda_c = \sqrt{\frac{\chi}{q' \Omega_\mathrm{P} \gamma}},
\label{lambda_c-eq}
\end{gather}
where $x_p$ is the distance from a planet to its corotation, $\lambda_c$  is the length scale over which a luminous planet releases heat, and $q' = 1.5$ is the shear parameter for a Keplerian disk.
The heating torque on the other hand is:
\begin{gather}
\gamma\frac{\Gamma_\mathrm{thermal}^\mathrm{heating}}{\Gamma_0} = 1.61 (\gamma -1) \frac{x_p}{\lambda_c} \frac{L}{L_c}
\\      L_c = \frac{4\pi G M \chi \rho_0}{\gamma},
\label{L_c-eq}
\end{gather}
where $L$ is the luminosity of the planet, $\rho_0$ is the unperturbed disk density and $L_c$ is used to normalize the density perturbation caused by the heat release.
See \autoref{L_crit-fig} for a plot of the critical luminosity at various times.
The luminosity here is the accretion luminosity of a planetary embryo, which, according to \cite{Chrenko} follows:
\begin{equation}
\label{accretion_luminosity-eq}
L = \frac{G M}{R} \dot{M},
\end{equation}
where $R$ is the radius of the planet and $\dot{M}$ is the solid accretion rate onto the planet.
The radius of the planet was determined from its mass by assuming the planet is spherical with a constant density of $5 \, \mathrm{\frac{g}{cm^3}}$, which corresponds to an Earth-like density.
Considering the density of ice giants, which is roughly $2 \, \mathrm{\frac{g}{cm^3}}$, we would get a smaller luminosity, meaning\ the heating torque would be reduced.
The sum of heating and cooling torque can be simplified to the final form:
\begin{gather}
\label{heating_torque-eq}
\Gamma_{\text {thermal}}^{\text {total}}=1.61 \frac{\gamma-1}{\gamma} \eta\left(\frac{H}{\lambda_{c}}\right)\left(\frac{L}{L_{c}}-1\right) \Gamma_{0}
,\\ \eta=\frac{\alpha}{3}+\frac{\beta+3}{6},
\end{gather}
where $H$ is the height of the disk.
As a consequence, when the planet's luminosity is smaller than the critical value from \autoref{L_c-eq}, the cooling torque dominates over the heating torque and the overall thermal torque is negative.

\section{Calculation of type II migration timescales}
\label{typeII_appendix}
Type II migration describes the movement of massive planets in the disk.
Massive here means that the planet opens a gap in the disk.
A gap corresponds to an area in which the surface density is smaller than 10\,\% of the unperturbed value.
The opening of a gap by the planet or synonymously the transition into type II migration is modeled following \cite{GapParameter}.
In particular, a gap parameter $g$ is calculated as:
\begin{equation}
g = \frac{3 H}{4 r_\mathrm{H}} + \frac{50 \nu H^2}{(q r_\mathrm{P})^2},
\label{gap_parameter-eq}
\end{equation}
where $r_{\mathrm{H}}=r\left[M_{\mathrm{c}} /\left(3 M_{\star}\right)\right]^{1 / 3}$ is the planet's Hill radius, that is,\:the radius in which the planet dominates accretion.
When the gap parameter is smaller than 3, the migration timescale gets calculated as a linear interpolation between type I and type II migration, similar as in \cite{dittkrist2014impacts}.
This corresponds to the hatched regions in the migration maps.
When this gap parameter is smaller than 1, the planet has opened a full gap and migrates completely in the type II regime, which is marked by cross hatched regions in the migration maps.
Larger values of the gap parameter indicate that the planet does not perturb the disk significantly and migrates completely in the type I fashion.
At later times, it is easier for a planet to open a gap because the disks aspect ratio decreases in time.

A recent study by \cite{kanagawa2018radial} suggests that the type II migration timescale gets calculated just like the type I migration timescale, but gets damped by the lower surface density in the gap.
The usual model, which is incorporated in this paper, however, suggests that the type II migration timescale is equal to the accretion time scale $\tau_\mathrm{visc}$ of the disk (e.g.,\ \citealt{durmann2015migration, robert2018toward}), with
\begin{equation}
\tau_\mathrm{visc} = r_\mathrm{P}^2 / \nu.
\end{equation}
Depending on the viscosity in the disk, the planetary migration rate changes.
In low-viscosity environments planets in type-II migration move slower and can thus accrete more material and grow bigger.
\cite{typeIITimescale} have studied how the planet can slow down type II migration even further when growing much more massive than the gas surrounding the gap, which slows down the viscous accretion.
When $M_\mathrm{P} > 4\pi \Sigma_\mathrm{g}  r_\mathrm{P}^2$, the migration gets slowed down, resulting in a formula for the type II migration timescale:
\begin{equation}
\tau_{\mathrm{II}}=\tau_\mathrm{visc} \cdot \max \left(1, \frac{M_{\mathrm{P}}}{4 \pi \Sigma_{\mathrm{g}} r_{\mathrm{P}}^{2}}\right).
\label{typeII_timescale-eq}
\end{equation}

\section{Additional Plots}
\begin{figure*}[!htb]
        \centering
        \begin{subfigure}{0.49\textwidth}
                \centering
                \includegraphics[width=\linewidth]{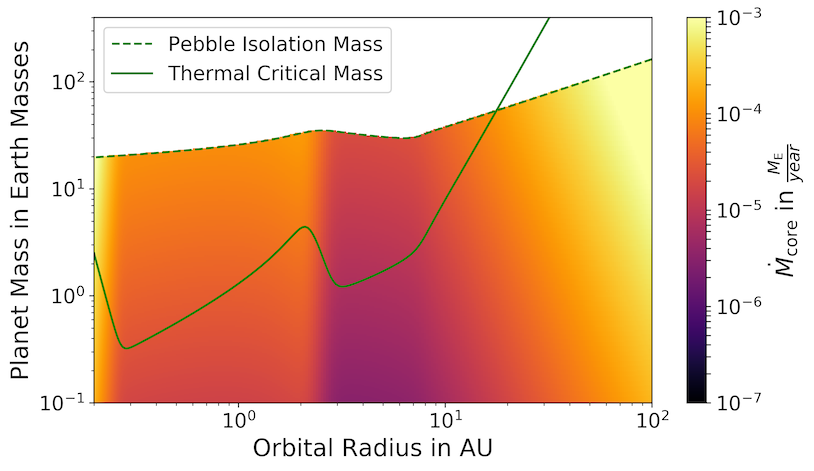}
                \caption{500\,kyrs}
        \end{subfigure}
        \begin{subfigure}{.49\textwidth}
                \centering
                \includegraphics[width=\linewidth]{./figures/_cont_2e-04_1000kyrs_JMHMdotminC.png}
                \caption{1\,Myr}
        \end{subfigure}
        
        \begin{subfigure}{0.49\textwidth}
                \centering
                \includegraphics[width=\linewidth]{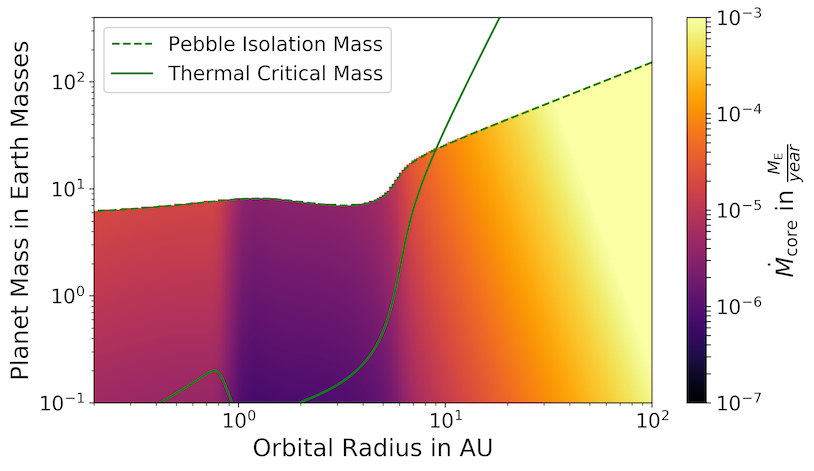}
                \caption{1.5\,Myr}
        \end{subfigure}
        \begin{subfigure}{.49\textwidth}
                \centering
                \includegraphics[width=\linewidth]{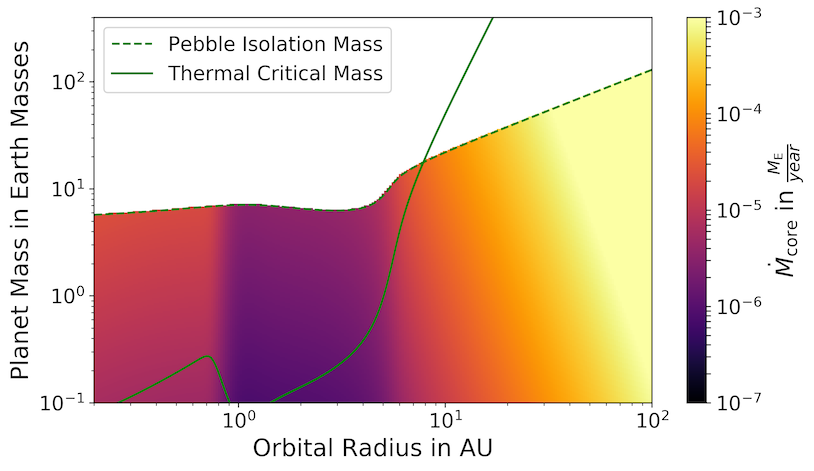}
                \caption{2\,Myr}
        \end{subfigure}
        \caption{Minimum core accretion rate that is required for the thermal torque to become positive in $M_\mathrm{E}$ / year for different times. It generally drops in time and is high in the inner and outer disk. At a few AU there is a sweet spot where the thermal torque can become positive more easily, caused by the critical luminosity (see \autoref{L_crit-fig}). For a more comprehensive review of these plots see \autoref{min_flux-fig}.
        }
        \label{mdotmin-fig_appendix}
\end{figure*}

\begin{figure*}[!htb]
        \centering
        \begin{subfigure}{0.49\textwidth}
                \centering
                \includegraphics[width=\linewidth]{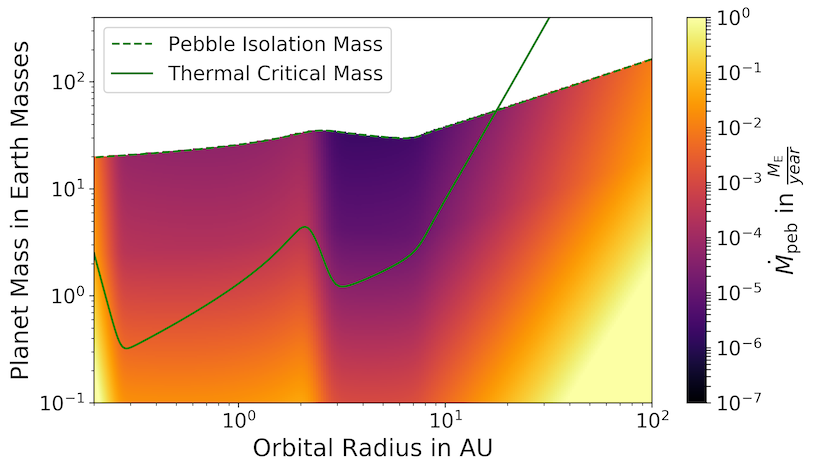}
                \caption{500\,kyrs}
        \end{subfigure}
        \begin{subfigure}{.49\textwidth}
                \centering
                \includegraphics[width=\linewidth]{./figures/_cont_2e-04_1000kyrs_JMHSigmaminC.png}
                \caption{1\,Myr}
        \end{subfigure}
        
        \begin{subfigure}{0.49\textwidth}
                \centering
                \includegraphics[width=\linewidth]{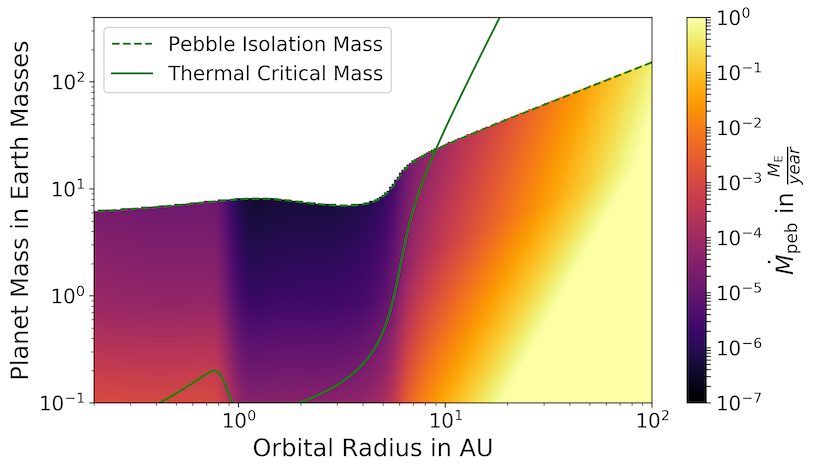}
                \caption{1.5\,Myr}
        \end{subfigure}
        \begin{subfigure}{.49\textwidth}
                \centering
                \includegraphics[width=\linewidth]{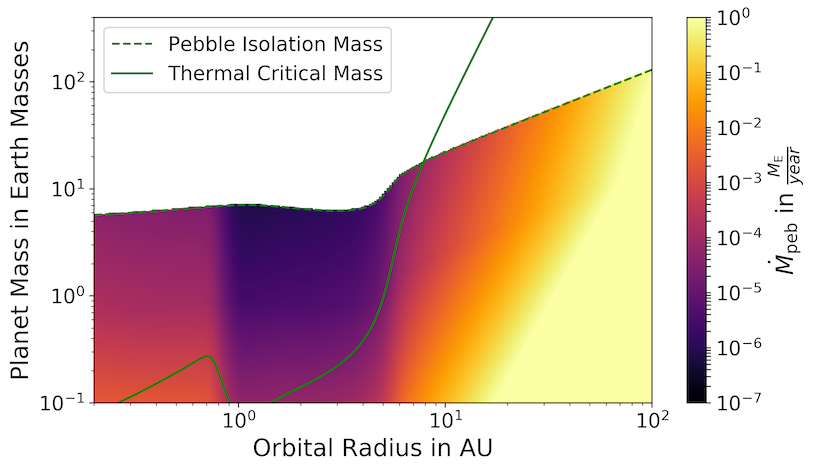}
                \caption{2\,Myr}
        \end{subfigure}
        \caption{Minimum pebble flux that is required for the thermal torque to become positive in $M_\mathrm{E}$ / year for different times. It generally drops in time and is high in the inner and outer disk. At a few AU there is a sweet spot, caused by the critical thermal diffusivity. As the thermal critical mass is only an estimate as of yet, values above are still plotted here. For a more comprehensive review of these plots see \autoref{min_flux-fig}.
        }
        \label{mdot_pebble_min-fig_appendix}
\end{figure*}

\begin{figure*}[!htb]
        \centering
        \begin{minipage}{\linewidth}
                \begin{subfigure}{.49\textwidth}
                        \centering
                        \includegraphics[width=\linewidth]{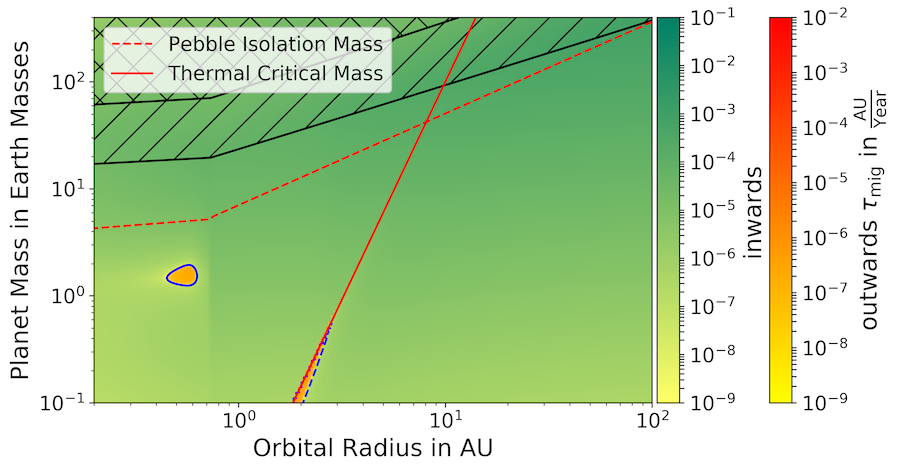}
                        \cite{Paardekooper}
                \end{subfigure}
                \begin{subfigure}{.49\textwidth}
                        \centering
                        \includegraphics[width=\linewidth]{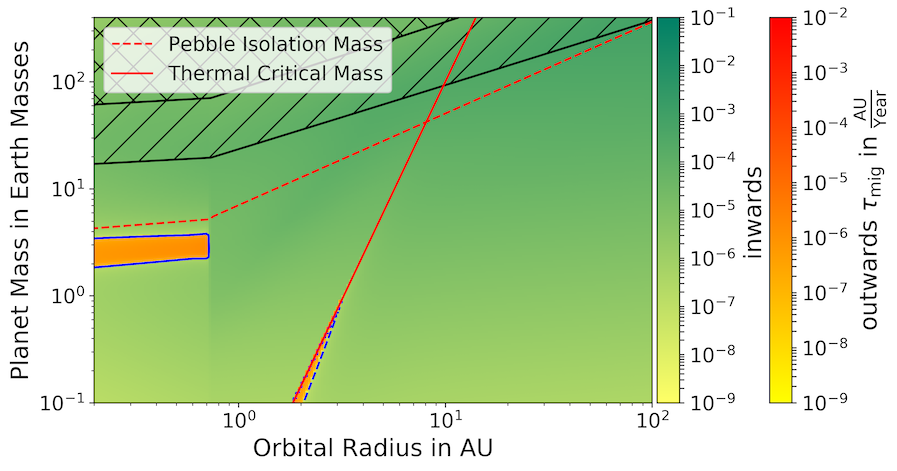}
                        \cite{ImprovedTypeI}
                \end{subfigure}
        \end{minipage}
        
        \caption{Migration maps in the \cite{IdaDisk} disk model after 1\,Myr at $\alpha = 5.4 \cdot 10^{-3}$.
                Like in \autoref{migrationmaps_visc-fig}, hatched regions mark the transition to type II migration, while planets in the cross-hatched areas are migrating completely in the type II regime and regions of outwards migration that are caused by the thermal torque are outlined by dashed blue lines, while regions of outwards migration inside solid blue lines are due to the original torque formulae.
                Compared to the corresponding plots in the \cite{DiscModel} disk (\autoref{migrationmaps_visc-fig_visc0-fig}), less outwards migration takes place.
                This is particularly true for the \cite{Paardekooper} formula and outwards migration due to the heating torque.}
        \label{Ida_migmaps}
\end{figure*}

\begin{figure*}[!htb]
        \centering
        \begin{minipage}{\linewidth}
                \begin{subfigure}{.49\textwidth}
                        \centering
                        \includegraphics[width=\linewidth]{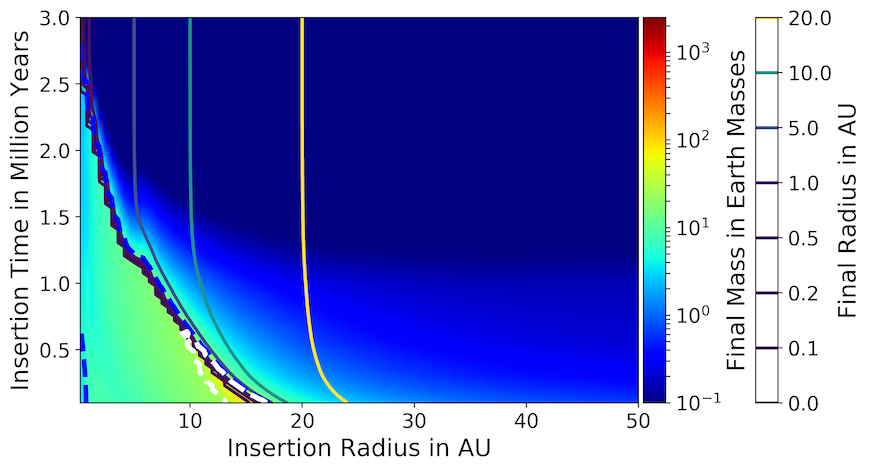}
                        \mbox{$\dot{M}_\mathrm{peb} = 2 \cdot 10^{-4} \cdot\exp (-\frac{t}{t_\mathrm{f}}) \, \frac{M_\mathrm{E}}{\mathrm{year}}$}, \cite{Paardekooper} + \cite{ThermalTorque}
                \end{subfigure}
                \begin{subfigure}{.49\textwidth}
                        \centering
                        \includegraphics[width=\linewidth]{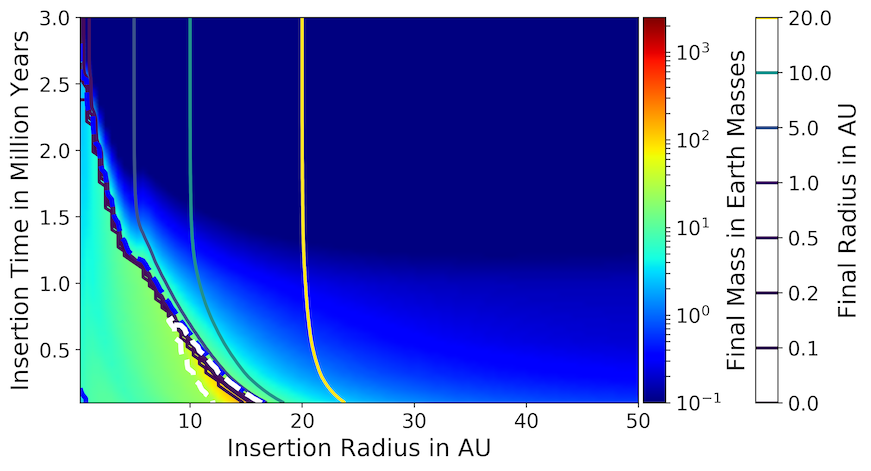}
                        \mbox{$\dot{M}_\mathrm{peb} = 2 \cdot 10^{-4} \cdot\exp (-\frac{t}{t_\mathrm{f}}) \, \frac{M_\mathrm{E}}{\mathrm{year}}$}, \cite{ImprovedTypeI} + \cite{ThermalTorque}
                \end{subfigure}
        \end{minipage}
        
        \caption{Planet evolution maps in the \cite{IdaDisk} disk model at $\alpha = 5.4 \cdot 10^{-3}$.
                As in \autoref{planet_evo-fig}, planets that are below the dashed blue line have reached pebble isolation mass and accrete gas.
                Planets that are interior to the dashed white line have reached runaway gas accretion and can grow very massive.
                The pebble flux is $\dot{M}_\mathrm{peb} = 2 \cdot 10^{-4} \cdot\exp (-\frac{t}{t_\mathrm{f}}) \, \frac{M_\mathrm{E}}{\mathrm{year}}$ in all plots.
                As in the \cite{DiscModel} disk, the formulae produce really similar results, however planets grow much lighter with the same pebble fluxes in the \cite{IdaDisk} model due to the region of outwards migration from corotation torques being smaller.}
        \label{Ida_evomaps}
\end{figure*}

\end{appendix}

\end{document}